\documentclass[aps,prb,amsmath,amssymb,twocolumn,superscriptaddress,nofootinbib]{revtex4-2}
\usepackage{amsmath,amssymb,amsthm,amsfonts,float,graphics,epsfig,epstopdf,color,verbatim,tabularx,bm,multirow,appendix,url}

\usepackage{footmisc}
\usepackage{endnotes}

\usepackage{graphicx}
\usepackage{bm}
\usepackage[colorlinks=true, citecolor=blue, linkcolor=blue, urlcolor=blue]{hyperref}
\usepackage{dsfont}
\usepackage{color}
\def\Tr{\mathop{\mathrm{Tr}}}
\usepackage{ulem}
\usepackage{xcolor}
\usepackage{dcolumn}
\usepackage{wasysym} 
\usepackage{xspace}
\usepackage{etoolbox}
\usepackage{orcidlink}

\newcommand{\ve}[1]{\boldsymbol{#1}}

\newcommand{\be}{\begin{equation}}
\newcommand{\ee}{\end{equation}}
\newcommand{\bea}{\begin{eqnarray}}
\newcommand{\eea}{\end{eqnarray}}

\newcommand{\sgn}{\text{sgn}}

\newcommand{\e}{e}

\begin{document}

\title{Dimensionality-Changing Transition from a Non-Fermi Liquid to a Spin-Solid in a Multichannel Kondo Lattice}
\author{Simon Martin} %
\thanks{These authors contributed equally to this work.}
\affiliation{Department of Physics, University of California at San Diego, La Jolla, California 92093, USA}
\author{Marcin Raczkowski \orcidlink{0000-0003-1248-2695}}%
\thanks{These authors contributed equally to this work.}
\affiliation{\mbox{Institut f\"ur Theoretische Physik und Astrophysik,
		Universit\"at W\"urzburg, 97074 W\"urzburg, Germany}}
\author{Fakher F. Assaad \orcidlink{0000-0002-3302-9243} }
\affiliation{\mbox{Institut f\"ur Theoretische Physik und Astrophysik,
		Universit\"at W\"urzburg, 97074 W\"urzburg, Germany}}
\affiliation{W\"urzburg-Dresden Cluster of Excellence ct.qmat, Am Hubland, 97074 W\"urzburg, Germany}
\author{Tarun Grover} %
\affiliation{Department of Physics, University of California at San Diego, La Jolla, California 92093, USA}

\begin{abstract}
	A multichannel Kondo system, where a single quantum spin couples to multiple channels of an electronic bath, provides one of the simplest examples of a zero-dimensional non-Fermi liquid. It is natural to ask: what happens when an extensive number of such systems are coupled together? A simple renormalization group argument implies that in a chain of SU$(N)$ multichannel quantum systems, where each spin is coupled to its own bath of $K$ channels, the individual spins dynamically decouple at low energy when $N > K$, resulting in a `sliding' non-Fermi liquid. Using Quantum Monte Carlo (QMC) simulations, we find evidences of a continuous, `dimensionality-changing' phase transition out of this non-Fermi liquid into a valence-bond solid phase as the intersite coupling is increased. Remarkably, at the critical point, correlations exhibit a power-law behavior even along the direction in which the spins are coupled, indicating the breakdown of dynamical decoupling at the transition. We also develop an RG scheme to understand the universal aspects of this transition.
\end{abstract}
\maketitle

\textbf{Introduction:} Zero-dimensional critical systems, that is,  quantum mechanical systems with power-law temporal correlations, have been a playground for a whole host of phenomena, ranging from physics of quantum impurities in a metal \cite{nozieres1980kondo,andrei1984solution,tsvelick1985exact,affleck1991the,affleck1991critical,affleck1993exact,ludwig1994exact,cox1993spin,parcollet1998overscreened,Cox98} to quantum chaos and holography \cite{bohigas1971two,brody1981random,sachdev1993gapless,georges2001quantum,Kitaev_KITP2015_talk,sachdev2015bekenstein,maldacena2016remarks}. It is natural to seek the fate of a collection of such systems as they couple with each other, a set-up that has attracted both theoretical \cite{gu2017local,song2017strongly,ben2018strange,chowdhury2018translationally,jian2017model,davison2017thermoelectric,khveshchenko2018thickening,komijani2018model,komijani2019emergent} and experimental  \cite{pouse2023quantum,karki2023z} attention. Analytical progress on this problem often involves taking a semiclassical limit where one considers sending the Hilbert space dimension of the zero-dimensional system to infinity before sending the total system volume to infinity. On the other hand, numerical progress with finite local Hilbert spaces is often stymied by a Monte Carlo sign problem. In this work we will study the fate of a lattice of  multichannel Kondo dots, each of which in isolation forms a 0+1-D non-Fermi liquid (NFL), and which are coupled  via a spin-exchange interaction. Motivated by renormalization group (RG) arguments,  we present an unbiased Quantum Monte Carlo (QMC) study of a class of such systems, and find that it exhibits a `dimensionality-changing' phase transition (DCPT) from a phase where different dots dynamically decouple (`a sliding non-Fermi liquid') to a valence-bond-solid ordered phase.

Let us briefly review the landscape of phenomena where the effective dimensionality changes with tuning parameters \cite{PhysRevLett.72.316,PhysRevLett.74.968,PhysRevLett.74.2997,Biermann01,PhysRevB.65.115117,Giamarchi04,PhysRevLett.99.126404,Marcin12,PhysRevB.84.045112,PhysRevLett.115.260401,Raczkowski15,PhysRevB.98.094403}. Consider a $d_{\parallel} + d_{\perp}$-space dimensional system built from $d_{\parallel}$-dimensional units weakly coupled along $\perp$ directions. For instance, Bechgaard salts  with $d_{\parallel} = 1, d_{\perp} = 2$  crossover to 3d  behavior only at low temperature~\cite{Vescoli98}. More intriguing is dynamical decoupling, where the inter-unit coupling $J_{\perp}$ becomes RG-irrelevant, enabling a genuine DCPT as $J_\perp$ increases. A central question is whether such transitions can be continuous and, if so, whether critical correlations become power-law both along the $\parallel$ and $\perp$ directions. In known examples, dynamical decoupling persists even at the critical point~\cite{emery2000quantum, vishwanath2001two,golubovic1998fluctuations,o1998sliding,wu2023two,lake2021subdimensional}. One of our motivations is to probe the possibility of DCPTs with power-law correlations even along the $\perp$ direction, thus indicating the emergence of extra dimensions in the critical regime.  We find that coupled multichannel Kondo systems provide a simple, controlled setting for this possibility.

\begin{figure}[h]
	\begin{center}
		\includegraphics*[width=0.62\columnwidth,angle=-0]{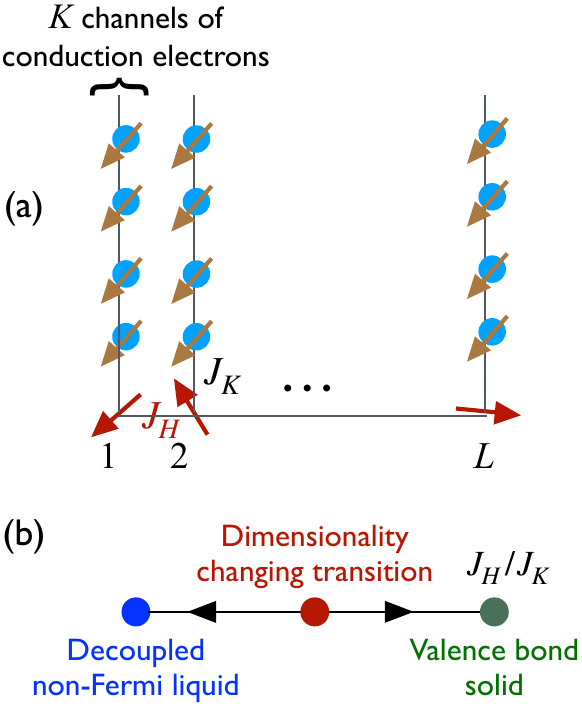} 
		\caption{(a) Geometry of the setup considered in this work: local moments carry an SU($N$) spin and form a 1d lattice, with each local moment coupled to separate $K$ channels of conduction electrons. (b) Phase diagram for $N = 4, K = 2$ based on sign-problem-free QMC simulations: as $J_H/J_K$ increases, one undergoes a phase transition from a phase where different local moments dynamically decouple and therefore the low-energy theory is a collection of decoupled 0+1-D non-Fermi liquids, to a phase where they strongly couple and form a valence-bond solid.}
		\label{Fig:phasedia}
	\end{center}
\end{figure}

The single-impurity multichannel Kondo problem is well understood: its ground state exhibits fractionalization and a 0+1-D NFL localized at the impurity~\cite{nozieres1980kondo,andrei1984solution,tsvelick1985exact,affleck1991the,affleck1991critical,affleck1993exact,ludwig1994exact,cox1993spin,parcollet1998overscreened,Cox98}, sharing features with the Sachdev-Ye-Kitaev (SYK) model~\cite{sachdev1993gapless,georges2001quantum,Kitaev_KITP2015_talk,sachdev2015bekenstein,maldacena2016remarks}, such as the presence of power-law correlations, finite ground-state entropy and quantum chaos~\cite{larzul2024fast}. Our interest is in possible DCPTs in lattices of $L$ multichannel Kondo dots made of SU$(N)$ spins coupled to their own set of $K$ channels of conduction electrons, see Fig.~\ref{Fig:phasedia} (a). We are interested in the limit $L \to \infty$, while keeping $N, K$ finite. Coupled 0+1-D non-Fermi liquids dots~\cite{gu2017local,song2017strongly,ben2018strange,chowdhury2018translationally,jian2017model,davison2017thermoelectric,khveshchenko2018thickening} are typically analyzed by first taking the semiclassical limit ($N \to \infty$) before sending $L \to \infty$. Since the order of limits $N \to \infty$, and $L \to \infty$ do not necessarily commute~\cite{mcgreevy2016tasi}, new methods are required. Two key observations allow us to make progress: first, RG implies the existence of a DCPT whenever $K < N$; we provide analytical results on this transition in a certain  regime of parameters. Secondly, we find that a large class of multichannel Kondo lattice problems do not suffer from the fermion sign problem. Exploiting this fact, and using extensive QMC simulations, we then study the case $K = 2, N = 4$, confirm the existence of a DCPT and characterize the phase diagram and the exotic critical point.

\textbf{Multichannel Kondo lattice:} Our model consists of a chain of multichannel Kondo dots (see Fig.~\ref{Fig:phasedia} (a)), where each dot has its own separate bath. In addition, the local moments are coupled to each other via a nearest-neighbor Heisenberg interaction $J_H$ (all repeated indices are summed over). The Hamiltonian is
\begin{eqnarray}
	\hat{H} & = & \sum_{x=1}^L \hat{H}_{\textrm{bath}}(x) 
	+ J_H \sum_{x=1}^L \hat{S}^{a}(x) \hat{S}^{a}(x+1) \nonumber \\
	& &  + J_K \sum_{x=1}^L  \hat{c}^{\dagger}_{i\alpha}(x,y=0) T^{a}_{\alpha \beta} \hat{c}_{i \beta}(x,y=0) \hat{S}^{a}(x) \, ,  \label{eq:MCK_lattice}
\end{eqnarray}
where $x$ denotes the location of the local moments. $\hat{H}_{\textrm{bath}}(x) = -t \sum_y (\hat{c}^{\dagger}(x,y) \hat{c}(x,y+1) + \textrm{h.c.})$ -- the precise form of $H_{\textrm{bath}}(x)$ is immaterial as long as it is local and has a non-zero local density of states. The conduction electrons carry two labels $i = 1, ..., K$ and $\alpha = 1, ..., N$ corresponding to the SU$(K)$ channel and SU$(N)$ spin degrees of freedom respectively. The label $a$ runs over the $N^2-1$ SU$(N)$ generators in the fundamental representation. One may represent the local moments as $\hat{S}^a = \hat{f}^{\dagger} T^a \hat{f}$ with the constraint $\hat{f}^{\dagger} \hat{f} = N/2$ so that the effective Hilbert space corresponds to the completely anti-symmetric (fermionic) representation of SU$(N)$ with $N/2$ rows and a single column. 

Our model  resembles that of Ref.~\cite{komijani2019emergent} with key differences. In \cite{komijani2019emergent}, the impurity transforms in a bosonic (instead of a fermionic) representation, and $K/N$ is chosen so that the single impurity's ground state is a Fermi liquid, whereas in our problem, it is a non-Fermi liquid. Moreover, when the Heisenberg term $J_H$ dominates, the bosonic representation leads to a translationally invariant ground state, while, as we will soon see, our model develops translational symmetry breaking spontaneously. Most importantly, Ref.~\cite{komijani2019emergent} focused on the (semiclassical) large-$N$ limit, and as discussed there, $1/N$ corrections can potentially induce magnetic instabilities of the Fermi liquid phase~\cite{lobos2012magnetic}. By contrast, our main motivation is to chart the phase diagram at finite $K, N = \mathcal{O}(1)$, using RG arguments and QMC simulations.  Analytical progress has also been made on few site coupled multichannel Kondo dots using BCFT techniques \cite{affleck1992exact,affleck1995conformal,georges1995solution, ingersent2005kondo}, but it is not clear how to extend this approach to a whole lattice worth of coupled multichannel dots.

\textbf{Renormalization group:}  For $J_H = 0$, the system reduces to decoupled 0+1-$D$ non-Fermi liquids, each described by the single-impurity multichannel Kondo problem. The asymptotic temporal correlation function for the local moments is $ \langle \hat{S}^{a}(x, \tau) \hat{S}^{a}(x, 0) \rangle  \sim 1/\tau^{2 \Delta^0_S}$ where $\Delta^{0}_S = 1/(1+K/N)$~\cite{nozieres1980kondo,andrei1984solution,tsvelick1985exact,affleck1991the,affleck1991critical,affleck1993exact,ludwig1994exact,cox1993spin,parcollet1998overscreened,Cox98}. Turning on a small Heisenberg coupling $J_H$, we note: (i) When $K <N$, using the aforementioned $\Delta^{0}_S$, $J_H$ is \textit{irrelevant} at the decoupled fixed point, implying a  stable \textit{dynamically decoupled}  phase at $0 < J_H \ll J_K$. (ii) Even if microscopically, only the nearest-neighbor interaction $J_H$ (Eq.~\ref{eq:MCK_lattice}) is introduced, RG will dynamically generate longer-ranged interactions of the form $\sum_{x, x'} J_x \hat{S}^{a}(x') \hat{S}^{a}(x'+x)$. At the decoupled fixed point, all couplings $J_x$ are equally (ir)relevant, and therefore, to carry out an RG analysis, one needs to keep all of them.  (iii) A controlled, perturbative RG calculation can be organized by introducing a small parameter $\epsilon = \frac{2}{1+K/N}-1$ that changes sign when $K = N$. Using conformal perturbation theory about the decoupled fixed point, one then finds the following $\beta$-functions (see Appendix \ref{Sec:Appendix_A} for more details):

\begin{align}
	\begin{split}
		\beta(J_x) &= -\epsilon J_x + \alpha J_x^2 - 2 J_{x/2}^2 \\ &\hspace{0.5cm} -4 \sum_{x'=1}^{\frac{L-1}{2}} \left(J_{x'} J_{x - x'} + J_{x'} J_{L - x - x'} + J_{x'} J_{x+x'}\right), \,
	\end{split} \label{eq:RG}
\end{align}
where for simplicity, $x$ is assumed to be even and $L$ is odd (see Appendix \ref{Sec:Appendix_A} for an equation valid for any $L,x$). The parameter  $\alpha > 0$ depends only on $N$ and $K$ \footnote{Explicitly, $\alpha = N C^2_{SSS}$, where $C_{SSS}$ is the operator product expansion coefficient related to the fusion of operators $\hat{S}^a \times \hat{S}^b \to \hat{S}^c$ in a single multichannel Kondo dot. See Appendix \ref{Sec:Appendix_A} for the calculation of $C_{SSS}$ in the large $N$ and $K$ limit.}.  We assume that the critical point is translationally invariant, and impose periodic boundary conditions along $x$. When $\epsilon > 0$ (i.e. $N > K$), the linear term in $J_x$ in Eq.~\ref{eq:RG} informs us that the dynamically decoupled phase is stable, yielding a `sliding non-Fermi liquid' (analogous to sliding Luttinger liquids~\cite{emery2000quantum, vishwanath2001two,golubovic1998fluctuations,o1998sliding}). 
In the opposite limit, when $J_H = J_{x=1} \gg J_K$ with $J_{x \neq 1} = 0$, the system reduces to an SU$(N)$ Heisenberg chain whose ground state is spontaneously dimerized for $N \geq 4$ ~\cite{affleck1985large,Affleck88,Arovas88,Read89}, and which is stable against Kondo coupling (see below). This establishes the existence of a DCPT for the Hamiltonian in Eq.~\ref{eq:MCK_lattice} whenever $N > K > 1$. Assuming that there are no intermediate, translation-symmetric phases, this leads to the phase diagram sketched in (Fig.~\ref{Fig:phasedia} (b)). Let us next discuss the critical properties within the aforementioned $\epsilon$-expansion. The coupled, non-linear RG Eqs.~\ref{eq:RG} admit a large class of solutions if one doesn't put any constraint on the starting point of the RG flow. However, we find that when $\alpha \gtrsim  8$, and if the starting point of the RG flow corresponds to a local Hamiltonian with short-range antiferromagnetic interactions (such as Eq.~\ref{eq:MCK_lattice}), then there always exists  a fixed point with a single relevant direction where all couplings $J^{*}_x$ are non-zero and decay exponentially with $x$: $|J^{*}_x| \sim e^{- x/\xi}$, with $\xi \sim 1/\log(\alpha)$, thereby describing the emergence of spatial locality. This is in notable contrast with DCPT in other contexts~\cite{emery2000quantum, vishwanath2001two,golubovic1998fluctuations,o1998sliding,wu2023two,lake2021subdimensional} where the dimensional reduction holds even at the critical point, i.e., the analog of these couplings vanishes at the critical point. At the fixed point, $J^{*}_1$ (i.e. nearest-neighbor interaction) is the dominant term, and has a positive sign (i.e. predominantly antiferromagnetic interactions). The fixed-point couplings $J^{*}_x$ determine the power-law decay exponent for temporal correlations via $\langle \hat{S}^{a}(x,0) \hat{S}^{a}(x,\tau) \rangle \sim 1/\tau^{2 \Delta_S}$, where $\Delta_S = \Delta^{0}_S + 4  \sum_{x'=1}^{\frac{L-1}{2}} J^{*}_{x'} $  (see Appendix \ref{Sec:Appendix_A}). On the other hand, when $\alpha = \mathcal{O}(1)$, we find that this fixed point is lost, and a new class of solutions with a single relevant direction appears, where $J^{*}_x$ oscillates in sign as a function of $x$ with $|J^{*}_x| \sim 1/L$. The envelope of $J^{*}_x$ does not  decay as a function of $x$, which implies that at any finite $L$, the equal-time, unequal space spin-spin correlation functions will have quasi long-range order of magnitude $1/L$. Intriguingly, such a fixed point can appear from an initial condition with short-range AFM interactions. We leave a detailed study of this exotic class of solutions to the future.

Let us also briefly discuss a peculiarity of the VBS phase that is expected at $J_H \gg J_K$. Due to the coupling to the conduction electrons, the spin-fluctuations in this phase, $\vec{\phi}$, acquire a Landau-damping term:  $S = \int d\omega dk\,\, |\vec{\phi}|^2 (k^2 + |\omega| + m^2)$, where $m$ is the gap to spinful excitations in the absence of the bath. Landau damping leads to power-law correlations along the time direction, $\langle \hat{S}^a(x,\tau) \hat{S}^a(x,0)\rangle \sim 1/\tau^2$, while the unequal-space, equal time correlations continue to decay exponentially as in an isolated one-dimensional chain with VBS order (due to the individual bath for each local moment, spatially long-ranged RKKY interactions are not induced). A large-$N$ mean-field calculation also arrives at the same results (see Appendix \ref{Sec:Appendix_C}).

\begin{figure}[h]
	\begin{center}
		\includegraphics*[width=0.95\columnwidth,angle=-0]{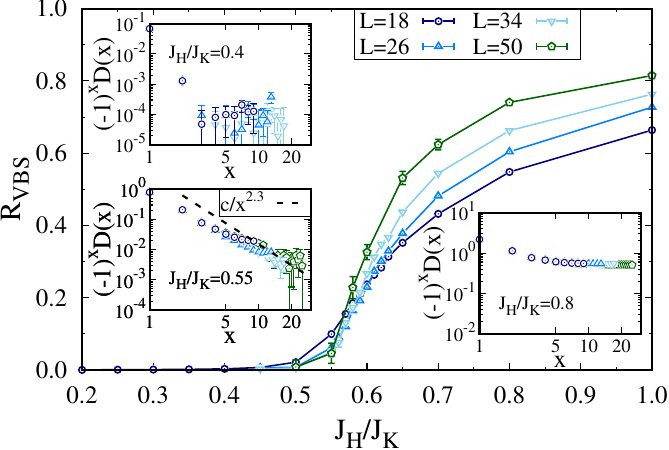}
		\caption{VBS correlation ratio $R_{\textrm{VBS}}$  for $N=4$, $K=2$ as a function of $J_H/J_K$  for $J_K=W=1$.
			Insets show the real space dependence of the dimer-dimer correlation function $(-1)^xD(x)$ at
			$J_H/J_K=0.4$ (decoupled phase), $J_H/J_K=0.55$ (DCPT), and $J_H/J_K=0.8$ (VBS phase).         
			At the DCPT, one observes a power-law decay of $(-1)^xD(x)$ with an exponent of 2.3. } 
		\label{Fig:R}
	\end{center}
\end{figure}

\textbf{QMC:}  Although the renormalization group analysis shows that there must exist at least two different phases when $N > K$, it is challenging to access unequal space critical correlations using it, since that would require solving the low-energy action at the RG fixed point describing the phase transition. RG also relies on a small parameter $\epsilon$ which is $\mathcal{O}(1)$ when $N, K$ are $\mathcal{O}(1)$. Therefore, in the rest of the paper, we will pursue a more direct approach that exploits the absence of sign problem in the microscopic Hamiltonian of Eq.~\ref{eq:MCK_lattice}. We will focus on $N = 4, K = 2$. We  have used the Algorithms for Lattice Fermions (ALF) \cite{ALF_v2} implementation of the  projective  auxiliary field quantum Monte Carlo algorithm \cite{Sorella89,Sugiyama86,AssaadEvertz2008}. A  key point to resolve low energy physics is to use a Wilson logarithmic  discretization~\cite{wilson1975renormalization}  of the electronic baths. We use a flat density of states of width $W$.
The logarithmic discretization makes the imaginary time projection, $\Theta$ challenging: our largest systems 
required  $\Theta J_K =250$. While the IR is faithfully captured, the UV is not. As a consequence, the value of the critical coupling at which we observe the DCPT depends on the step of the logarithmic discretization $\Lambda$. The trial wave function is chosen so as to respect symmetries of the   model. This includes SU$(K)$ channel and SU$(N)$ spin symmetries as well as the U(1) charge symmetry accounting for charge conservation in each bath. We find that none of these symmetries are  spontaneously broken in our simulations.
A  detailed implementation of the algorithm can be found in Appendix \ref{Sec:Appendix_B}.

To determine the critical value of  $J_H^c/J_K^{}$ at which  dimer order develops, we calculate the dimer structure factor for the SU($N$) spins:
$D(q)    =
\frac{1}{L} \sum_{x,x'}  e^{i q  \left( x - x' \right) }  D(x-x')  $
with $D(x-x') = \langle \hat{\Delta}_{x,x+1} \hat{\Delta}_{x',x'+1} \rangle - \langle \hat{\Delta}_{x,x+1} \rangle \langle \hat{\Delta}_{x',x'+1} \rangle  $ and  $\hat{\Delta}_{x,x+1} = \hat{S}^{a}(x)\hat{S}^{a}(x+1)$. We use it to construct the correlation ratio $R_{\textrm{VBS}} =  1 - \frac{D^{}(Q - \delta q) }{D^{}(Q) } $, where $Q=\pi$ is the ordering wave vector and $\delta q=2\pi/L$ is the smallest finite wave vector
for a given system size $L$. As a renormalization group invariant quantity, $R_{\textrm{VBS}}$ shall scale to unity (zero) in the ordered (disordered) phase, respectively, 
and we notice in Fig.~\ref{Fig:R}  that  $R_{\textrm{VBS}}\to 0$  at the critical point  $J_H^c/J_K^{}\simeq 0.55$.  
In one dimension, if correlation functions fall off quicker than $1/x$, then $D(q)$  is a continuous function of $q$ and the 
correlation ratio vanishes. The inset at criticality confirms this: the scaling  dimension of the  dimer operator is given by $\Delta_D \simeq  1.15$. The insets of Fig.~\ref{Fig:R}  show that below (above) $J_H^c $  the dimer correlations decay exponentially (converge to a constant). 
This supports a DCPT.

\begin{figure}[h!]
	\begin{center}
		\includegraphics*[width=0.48\columnwidth,angle=-0]{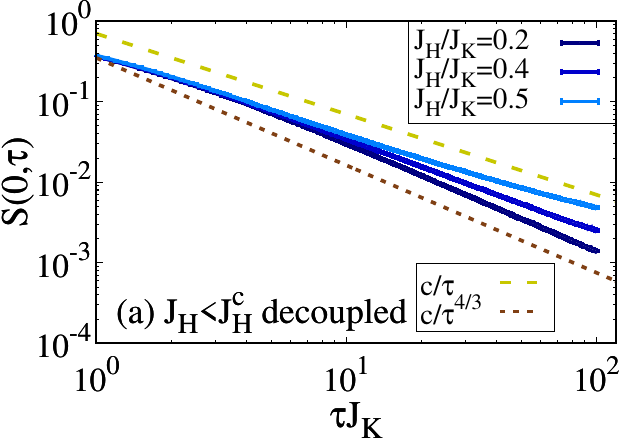}
		\includegraphics*[width=0.48\columnwidth,angle=-0]{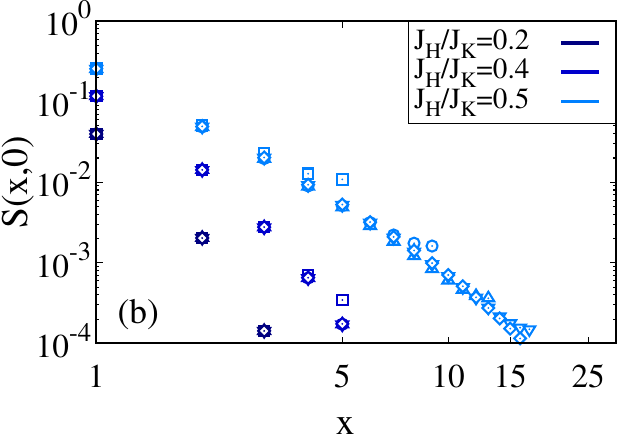}\\
		\includegraphics*[width=0.48\columnwidth,angle=-0]{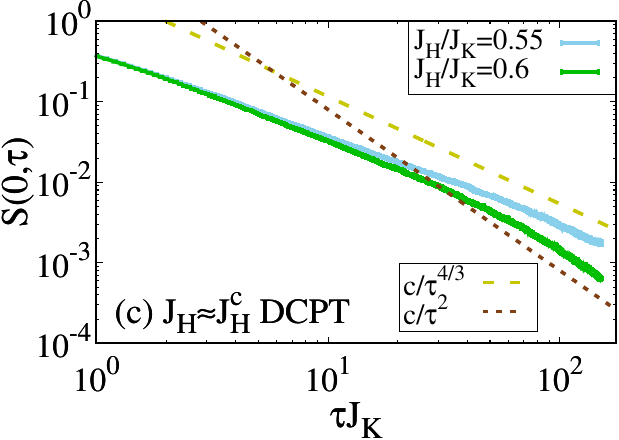}
		\includegraphics*[width=0.48\columnwidth,angle=-0]{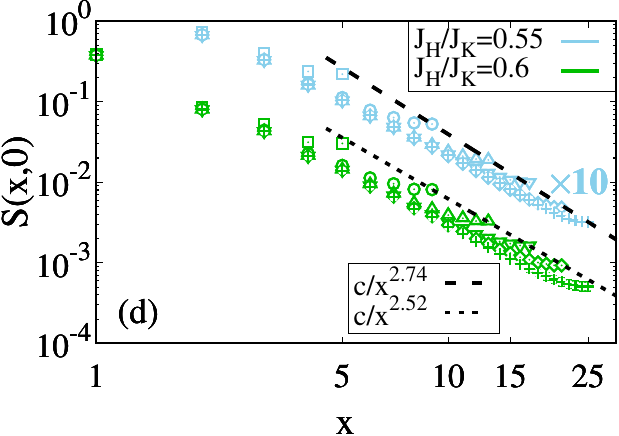}\\
		\includegraphics*[width=0.48\columnwidth,angle=-0]{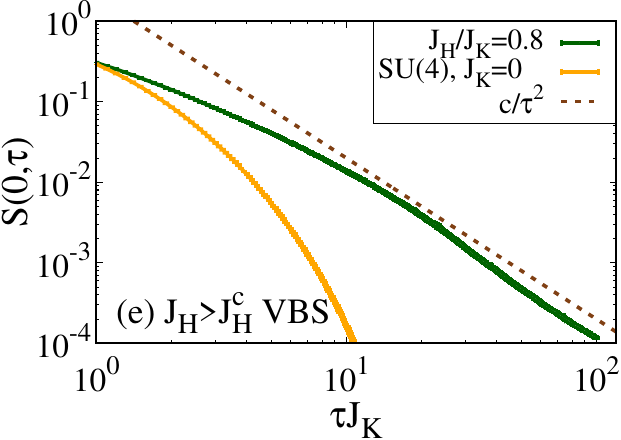}
		\includegraphics*[width=0.48\columnwidth,angle=-0]{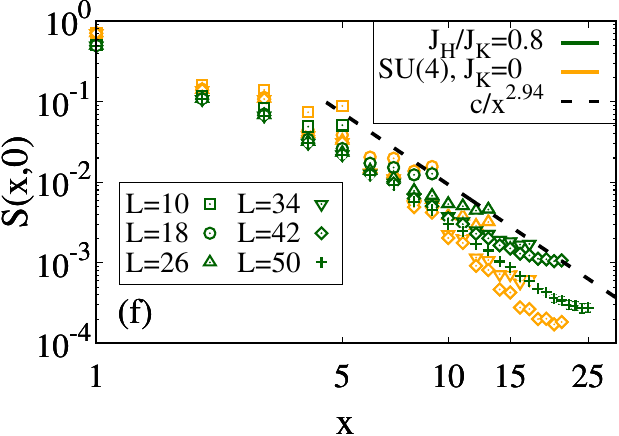}
		\caption{Space and time spin-spin correlation functions $S(x,\tau)$ at $N=4$ and $K=2$
			for different  $J_H/J_K$, corresponding to: (a,b) decoupled phase, (c,d) DCPT, and  (e,f) VBS phase.
			For comparison, in panels (e) and (f), we also plot the data for the decoupled ($J_K=0$) SU(4) Heisenberg chain with
			a clear exponential decay of $S(x,\tau)$.  
		}
		\label{Fig:S}
	\end{center}
\end{figure}

We now proceed to provide evidences of the dimensionality-changing nature of the phase transition.
To this end, we plot in Fig.~\ref{Fig:S} space and time spin-spin correlation functions 
$S(x,\tau)= e^{iQx} \langle \hat{S}^{a}(x, \tau)\hat{S}^{a}(0, 0) \rangle$
and discuss them from the point of view of the quantum critical fan:  a high temperature region or equivalently
short range/time correlations should reveal the quantum critical  scaling while the long distance/time limit
shall be dominated by the ground state properties.
As can be seen in Fig.~\ref{Fig:S} (a), the long time tail of $S(0,\tau)$ continues to display
the $K=2$ channels SU(4) Kondo impurity asymptotic behavior
$1/\tau^{\frac{2N}{N+K}}=1/\tau^{\frac{4}{3}}$ at $J_H/J_K=0.2$ and starts to weakly deviate only at $J_H/J_K=0.4$.
Within this range of $J_H/J_K$,  $S(x,0)$  falls off exponentially,  see Fig.~\ref{Fig:S} (b).
Upon further increasing $J_H/J_K$, the short range part of $S(x,0)$ progressively acquires a power law  whose exponent reaches its minimum in the vicinity of DCPT, see  Figs.~\ref{Fig:S} (b) and \ref{Fig:S} (d). 
In this case, the algebraic decay is observed even for our longest chains with $L=50$ sites.  
An intriguing aspect is that as one approaches the critical point from the decoupled phase, one observes a non-monotonic behavior of the temporal correlations $S(0,\tau)$, 
see Figs.~\ref{Fig:S} (a) and \ref{Fig:S} (c). 
Finally, at $J_H/J_K=0.8$, corresponding to the VBS phase (Fig.~\ref{Fig:S} (e)), a long time tail of $S(0,\tau)$ falls
off as $1/\tau^2$ in agreement with the aforementioned low energy theory of Landau damping and large-$N$ mean-field calculation. 
Remarkably, the corresponding real space correlations $S(x,0)$ measured on chains 
with up to $L=42$ sites continue to display a power law decay despite the coexisting long range VBS order. 
That stands in sharp contrast to a decoupled ($J_K=0$) dimerized SU(4) Heisenberg chain with a clear exponential decay 
of $S(x,0)$, see Figs. ~\ref{Fig:S} (e) and \ref{Fig:S} (f).
However, this is merely a signature of a crossover phenomenon and we were able to confirm that the expected  exponential 
fall-off becomes apparent in longer chains with $L\ge 50$.
We have equally checked that the bath spin-spin correlations $S_c(x,0)$  along the chain are non-vanishing and 
for $J_H^{}/J_H^c\ge 1$ they decay  within the errorbars  with the same exponents as those of the  local SU(4)  spin  degrees 
of freedom (see Appendix \ref{Sec:Appendix_B}).

\begin{figure}[h]
	\begin{center}
		\includegraphics[width=0.48\columnwidth,angle=-0]{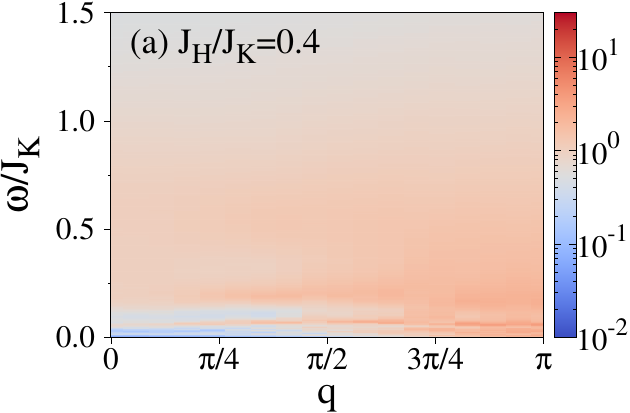}
		\includegraphics[width=0.48\columnwidth,angle=-0]{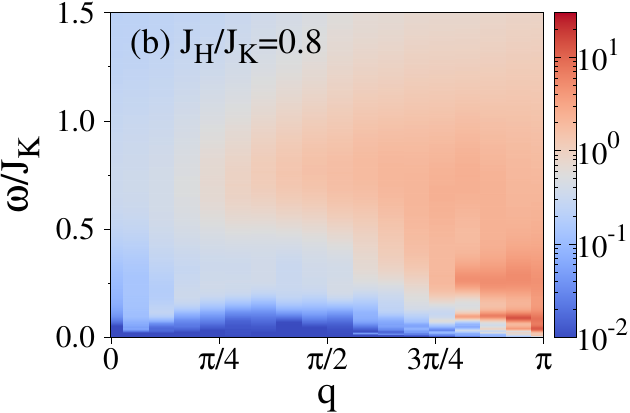}
		\caption{Dynamical spin structure factor  $S(q,\omega)$ in:
			(a) decoupled phase at $J_H/J_K=0.4$ and (b) VBS phase at $J_H/J_K=0.8$. }
		\label{fig:Sqw_main}
	\end{center}
\end{figure}

The dimensionality-changing transition has clear signatures in  the spin dynamics $S(q,\omega)$ extracted here using 
the  ALF library~\cite{ALF_v2}  implementation of the stochastic analytical continuation methods~\cite{beach04identifying,sandvik1998stochastic} 
from the QMC time displaced spin-spin correlation function. In particular, as can be seen in Fig.~\ref{fig:Sqw_main}, 
a finite dispersion of the excitation spectrum is generated across the transition as one exits the dynamically decoupled phase. Notably, due to the bath, spin excitations in the VBS phase remain gapless. 

\textbf{Discussion:} Perhaps the most interesting feature of our model is that it exhibits a continuous, dimensionality-changing transition where both spatial and temporal correlations decay as a power-law. Unlike coupled SYK models, where a direct hopping between SYK-dots destroys the decoupled non-Fermi liquid and results in a Fermi liquid at low temperature~\cite{song2017strongly}, in our model, the decoupled phase is stable as long as the symmetries of the model are maintained. Previous works have also studied dimensional reduction and associated dimensionality-changing phase transitions in various contexts, but to the best of our knowledge, in all such examples, one finds that the spatial correlations decay exponentially along the direction where the lower-dimensional systems are being coupled~\cite{emery2000quantum, vishwanath2001two,golubovic1998fluctuations,o1998sliding,wu2023two,lake2021subdimensional}. A full RG treatment of the critical spatial correlations in our problem remains an interesting open question. We also note that the RG prediction for the critical temporal correlations at the leading order in the $\epsilon$ expansion differs qualitatively from the QMC results -- the RG predicts that scaling dimension is greater than that in the decoupled phase, while the QMC suggests it is smaller. This suggests that the higher order contributions in the $\epsilon$-expansion may be important. Although we focused on multichannel Kondo dots coupled in a specific way, our methodology of employing Wilson numerical RG with QMC opens the window to explore a large class of open problems that have been discussed in the literature and where there remains several unanswered questions. For example, it allows one to explore a setup where the non-Fermi liquid behavior of a 0+1-D system gets imprinted on a larger dimensional system through coupling with an array of such 0+1-D systems~\cite{ben2018strange}. Another problem worth mentioning is that of multichannel Kondo dots coupled \textit{ferromagnetically}. Our RG analysis implies that in such a setup, when $K > N$, decoupled dots becomes immediately unstable towards a new, critical \textit{phase}. Another point worth investigating more is the landscape of phase transitions encoded in our RG. As mentioned earlier, the RG admits solutions where the couplings are long-ranged with oscillating signs, which is rather intriguing. Since RG is controlled by the small parameter $\epsilon = \frac{2}{1+K/N}-1$, we anticipate that these solutions will manifest themselves for sufficiently small $\epsilon$ (e.g. $N = 10, K= 8$), and it will be interesting to explore this possibility using QMC. Given the richness of the phase diagram of such coupled multichannel systems, we hope that our results will be a motivation to push beyond experiments such as Refs. \cite{pouse2023quantum,karki2023z} that study a few sites coupled multichannel Kondo systems.

\textbf{Acknowledgments}
We thank John McGreevy for helpful discussions. The authors gratefully acknowledge the Gauss Centre for Supercomputing e.V. (www.gauss-centre.eu) for funding this project by providing computing time on the GCS Supercomputer SUPERMUC-NG at Leibniz Supercomputing Centre
(www.lrz.de) as well as through the John von Neumann Institute for Computing (NIC) on the GCS Supercomputer JUWELS  at the  J\"ulich Supercomputing Centre (JSC).  We  also gratefully acknowledge the scientific support and HPC resources provided by  the Erlangen National High Performance Computing Center (NHR@FAU) of the Friedrich-Alexander-Universität Erlangen-Nürnberg (FAU) under NHR project 80069 provided by federal and Bavarian state authorities. NHR@FAU hardware is partially funded by the German Research Foundation (DFG) through grant 440719683. 
F.F.A. acknowledges financial support from the German Research Foundation (DFG) under the grant AS 120/16-1 (Project number 493886309) that is part of the collaborative research project SFB QMS funded by the Austrian Science Fund (FWF) F 86.
M. R.  acknowledges funding from the Deutsche Forschungsgemeinschaft (DFG, German Research Foundation), Project No. 332790403 as well as the W\"urzburg-Dresden Cluster of
Excellence on Complexity and Topology in Quantum Matter ct.qmat (EXC 2147, Project ID 390858490). T.G. is supported by
the National Science Foundation under Grant No. DMR-2521369.

\onecolumngrid 
\newpage
\appendix

\section{RG calculation}\label{Sec:Appendix_A}

This appendix provides details of the RG calculation for the coupled SU$(N)$ multichannel Kondo models using conformal perturbation theory. The Hamiltonian governing our system is

\begin{align}\label{Eq:Hamiltonian}
	\begin{split}
		\hat{H} &=  \sum_{i=1}^{K} \sum_{\alpha=1}^{N}
		\sum_k \sum_{x=1}^L \epsilon(k) \hat{c}^{\dag}_{i \alpha}(x,k) \hat{c}_{i \alpha}(x,k) + J_K \sum_{i=1}^K \sum_{\alpha,\beta=1}^N \sum_{a=1}^{N^2-1} \sum_{x=1}^L \hat{S}^a(x) \hat{c}^{\dag}_{i\alpha}(x,y=0) T^{a}_{\alpha \beta} \hat{c}_{i\beta}(x,y=0) \\ &\hspace{0.5cm} + J_H \sum_{a=1}^{N^2-1} \sum_{x=1}^L \hat{S}^a(x) \hat{S}^a(x+1) \, ,
	\end{split}
\end{align}

\noindent where the first term describes the conduction electron baths, with $k$ the conduction electron momentum, $i=1,...,K$ labels the channel, $\alpha=1,...,N$ labels the spin and $x=1,...,L$ labels the impurity site. We use periodic boundary conditions along $x$. The second term describes the Kondo interaction (at the origin of the bath $y=0$) between the conduction electrons and an SU$(N)$ impurity $\hat{S}^a$, transforming in the fully anti-symmetric representation with $N/2$ rows and a single column. The impurity is written in terms of fermionic spinons: $\hat{S}^a = \hat{f}_{\alpha}^{\dag} T_{\alpha \beta}^a \hat{f}_{\beta}$, where $\hat{f}_{\alpha}$ obeys the constraint $\hat{f}_{\alpha}^{\dag} \hat{f}_{\alpha} = N/2$ on each site and the $T^a$'s are the $N^2-1$ generators of SU$(N)$ in the fundamental representation. The last term corresponds to the nearest-neighbor Heisenberg interaction between the impurities.

\subsection{Decoupled fixed point}

Let us first consider the situation where $J_H=0$, which corresponds to a set of $L$ decoupled dots. The single-impurity multichannel Kondo model has been extensively studied in the literature \cite{nozieres1980kondo,parcollet1998overscreened,affleck1991the,affleck1991critical,affleck1993exact,ludwig1994exact} and has been shown to flow in the IR to a 1+1D SU$(N)_K$ Wess-Zumino-Witten (WZW) boundary conformal field theory (bCFT) for $K>1$ (for the fully-antisymmetric representation). This intermediate fixed point describes a 0+1D non-Fermi liquid. At this fixed point, the impurity $S^a$ acquires the scaling dimension of the primary operator of the SU$(N)_K$ algebra associated with the $(N^2-1)$-dimensional adjoint representation of SU$(N)$, denoted $\phi^a$, with scaling dimension $\Delta_{\phi} = \frac{N}{N+K} = \frac{1}{1+\gamma} \equiv \Delta_S^{0}$, with $\gamma = \frac{K}{N}$. This comes from the fact that $\phi^a$ is the most relevant boundary operator with the same symmetries as $S^a$ (the next operator is the current $J^a$ which has dimension $\Delta_J=1$ and is thus marginal on the boundary). Therefore, we make the replacement $S^a \rightarrow \phi^a$ in the IR, despite the fact that we will still use the symbol $S^a$.

The RG calculation will require the knowledge of the OPE of $S^a$ with itself in the single-impurity bCFT

\begin{equation} \label{Eq:OPE_BCFT}
	S_x^a(\tau) S_x^b(0) = C_{SS\mathds{1}} \frac{\delta^{ab}}{|\tau|^{2\Delta_S^0}} \mathds{1}_x + i C_{SSS} \sgn(\tau) \frac{f^{abc}}{|\tau|^{\Delta_S^0}} S^c_x(0) + ... \, .
\end{equation}

\noindent This form can be justified from the invariance under the anti-unitary ``time-reversal'' symmetry: $f_{\alpha} \rightarrow f_{\alpha}^{\dag} \implies S^a \rightarrow -S^a$, $i \rightarrow -i$. The ellipsis denotes more irrelevant terms, while subscripts $x$ emphasize that each impurity has its own OPE. $C_{SS\mathds{1}}$ and $C_{SSS}$ are the OPE coefficients associated respectively with the fusions $\phi \times \phi \rightarrow \mathds{1}$ and $\phi \times \phi \rightarrow \phi$. In what follows, we work with a normalized impurity 2-point function $\langle S^a(\tau) S^b(0) \rangle = \delta^{ab}/|\tau|^{2\Delta_S^0}$, so that we choose $C_{SS\mathds{1}} = 1$. $f^{abc}$ is the fully antisymmetric structure constant of SU$(N)$.

\subsection{Conformal perturbation theory}

We can now perform the RG analysis by using conformal perturbation theory. The decoupled IR bCFTs described by the critical action $S^* = \sum_{x=1}^L S^*_{x}$ are perturbed by Heisenberg interactions. As we will soon show it, one cannot only consider the nearest-neighbor interaction, as interactions with larger separations (next-nearest-neighbor,...) will also be generated. Let us start by defining the operator

\begin{equation}\label{Eq:O_x}
	O_x(\tau) = \sum_{x'=1}^{L} S_{x'}^a(\tau) S_{x'+x}^a(\tau) \, .
\end{equation}

\noindent For simplicity, we will start by detailing the case of an odd number of impurities $L$ (results for $L$ even are presented below). One can easily show that $O_{L-x}=O_x$, which means that we only have to consider operators $O_x$ in the interval $x=[1,\frac{L-1}{2}]$. Therefore, the full action is $S=S^*+\delta S$, where the perturbations are

\begin{equation} \label{Eq:deltaS}
	\delta S = \sum_{x=1}^{\frac{L-1}{2}} J_x \int d\tau O_x(\tau) + h \int d\tau O_h(\tau) \, ,
\end{equation}

\noindent where the last term, which contains the operator $O_h(\tau) = \sum_{x=1}^L \sum_a S_x^a(\tau)$, plays the role of a magnetic field and will be useful to compute the scaling dimension of $S^a_x$. The RG flow equations for the couplings $J_x$ and $h$ are obtained at quadratic order using the general formula \cite{cardy1996scaling}

\begin{equation}\label{Eq:Cardy}
	\beta(g_k) = (D-\Delta_k) g_k - \sum_{i,j} C_{ijk} g_i g_j + ... \, ,
\end{equation}

\noindent where $g_i$ represents the couplings $J_x$ and $h$, $\Delta_k$ is the scaling dimension of operator $O_k$ and $C_{ijk}$ are the OPE coefficients of the fusion $O_i \times O_j \rightarrow O_k$ for the operators of Eq.~\ref{Eq:deltaS}.

Let us emphasize that the unperturbed bCFTs  live in $D=0+1$ spacetime dimension and thus do not know about the spatial dimension along the spin chain. Since $O_x$ couples two different CFTs (at two different sites), its scaling dimension is simply $2\Delta_S^{0}$ and the RG equation for $J_x$ at tree-level is thus

\begin{equation}
	\beta(J_x) = \frac{d J_x}{dl} = (1 - 2 \Delta_S^{0}) J_x + \mathcal{O}(J_x^2) = \Big[ 1 - \frac{2}{1+\gamma} \Big] J_x + \mathcal{O}(J_x^2) = -\epsilon J_x + \mathcal{O}(J_x^2) \, ,
\end{equation}

\noindent where $\epsilon = \frac{2}{1+\gamma}-1$. Hence, for $\gamma \approx 1$, we have $|\epsilon| \ll 1$, which can be used to control the RG. Even if $\gamma$ is the ratio of two integers, it is treated as a continuous parameter for the sake of the RG calculation. Moreover, note that the sign of $\epsilon$ depends on whether $\gamma < 1$ ($K<N$), for which $\epsilon > 0$ or $\gamma > 1$ ($K>N$), for which $\epsilon<0$.

We note that the idea of obtaining RG equations for a stack of conformally invariant systems (albeit in a different context) has also been discussed in Refs. \cite{lake2021subdimensional,lake2021stability}.

\subsubsection{OPE coefficients of perturbing operators}

Having established all of this, the goal is now to compute the OPE coefficients $C_{ijk}$. Let us start with the OPE of $O_x$ with itself

\begin{align}
	\begin{split}
		O_x(\tau) O_x(0) &=  \sum_{x',x''=1}^L S_{x'}^a(\tau) S_{x'+x}^a(\tau) S_{x''}^b(0) S_{x''+x}^b(0) \\ &= \sum_{x'=1}^L \Big[ S_{x'}^a(\tau) S_{x'}^b(0) \Big] \Big[ S_{x'+x}^a(\tau) S_{x'+x}^b(0) \Big] + \sum_{x'=1}^L \Big[ S_{x'+x}^a(\tau) S_{x'+x}^b(0) \Big] S_{x'}^a(\tau) S_{x'+2x}^b(0) \\ &\hspace{1cm} + \sum_{x'=1}^L \Big[ S_{x'}^a(\tau) S_{x'}^b(0) \Big] S_{x'+x}^a(\tau) S_{x'-x}^b(0) + ... \, ,
	\end{split}
\end{align}

\noindent where the three terms correspond respectively to $x''=x'$, $x''=x'+x$ and $x''=x'-x$, while ellipsis denote terms with four spins on different sites. We now apply the single-impurity OPE to the brackets

\begin{align}
	\begin{split}
		O_x(\tau) O_x(0) &= \sum_{x'=1}^L \Bigg[ \frac{\delta^{ab}}{|\tau|^{2\Delta_S^{0}}} \mathds{1}_{x'} + i C_{SSS} \sgn(\tau) \frac{f^{abc}}{|\tau|^{\Delta_S^{0}}} S^c_{x'}(0) \Bigg] \Bigg[ \frac{\delta^{ab}}{|\tau|^{2\Delta_S^{0}}} \mathds{1}_{x'+x} + i C_{SSS} \sgn(\tau) \frac{f^{abd}}{|\tau|^{\Delta_S^{0}}} S^{d}_{x'+x}(0) \Bigg] \\ &\hspace{0.7cm} +\sum_{x'=1}^L \Bigg[ \frac{\delta^{ab}}{|\tau|^{2\Delta_S^{0}}} \mathds{1}_{x'+x} + i C_{SSS} \sgn(\tau) \frac{f^{abd}}{|\tau|^{\Delta_S^{0}}} S^d_{x'+x}(0) \Bigg] S_{x'}^a(\tau) S_{x'+2x}^b(0) \\ &\hspace{0.7cm} +\sum_{x'=1}^L \Bigg[ \frac{\delta^{ab}}{|\tau|^{2\Delta_S^{0}}} \mathds{1}_{x'} + i C_{SSS} \sgn(\tau) \frac{f^{abd}}{|\tau|^{\Delta_S^{0}}} S^d_{x'}(0) \Bigg] S_{x'+x}^a(\tau) S_{x'-x}^b(0) + ... \\ &= \frac{(N^2-1)}{|\tau|^{4\Delta_S^{0}}} \sum_{x'=1}^L \mathds{1}_{x'} \mathds{1}_{x'+x} - C_{SSS}^2 \frac{N}{|\tau|^{2\Delta_S^{0}}} \sum_{x'=1}^L S_{x'}^a(0) S^a_{x'+x}(0) \\ &\hspace{0.7cm} + \frac{1}{|\tau|^{2\Delta_S^{0}}} \sum_{x'=1}^L S_{x'}^a(\tau) \mathds{1}_{x'+x} S_{x'+2x}^a(0) + \frac{1}{|\tau|^{2\Delta_S^{0}}} \sum_{x'=1}^L  \mathds{1}_{x'} S_{x'+x}^a(\tau) S_{x'-x}^a(0) + ... \\ &\approx \frac{N^2-1}{|\tau|^{4\Delta_S^{0}}} \mathds{1} - \frac{N C_{SSS}^2}{|\tau|^{2\Delta_S^{0}}} O_x(0) + \frac{2}{|\tau|^{2\Delta_S^{0}}} O_{2x}(0) + ... \, ,
	\end{split}
\end{align}

\noindent where we have used $\delta^{ab} \delta^{ab} = N^2-1$ and $f^{abc} f^{abd} = N \delta^{cd}$. The last equality follows from Taylor expanding to leading order in $\tau$ and using periodic boundary conditions along $x$. Ellipses denote terms with three or more spins, which are irrelevant operators. From this, we identify the following two OPE coefficients:

\begin{equation}
	C_{xxx} = - N C_{SSS}^2 \, , \quad C_{xx,2x} = 2 \, .
\end{equation}

We now move on to the OPE of $O_x$ with $O_{x'}$, with $x' \neq x$

\begin{align}
	\begin{split}
		&O_x(\tau) O_{x'}(0) \\ &= \sum_{x'',x'''=1}^L S_{x''}^a(\tau) S^a_{x''+x}(\tau) S_{x'''}^b(0) S_{x'''+x'}^b(0) \\ &= \sum_{x''=1}^L \Big[ S_{x''}^a(\tau) S_{x''}^b(0) \Big] S^a_{x''+x}(\tau) S_{x''+x'}^b(0) + \sum_{x''=1}^L \Big[ S_{x''}^a(\tau) S_{x''}^b(0) \Big] S^a_{x''+x}(\tau) S_{x''-x'}^b(0) \\ &\hspace{0.5cm}+ \sum_{x''=1}^L \Big[ S_{x''+x}^a(\tau) S_{x''+x}^b(0) \Big] S^a_{x''}(\tau) S_{x''+x+x'}^b(0) + \sum_{x''=1}^L \Big[ S_{x''+x}^a(\tau) S_{x''+x}^b(0) \Big] S^a_{x''}(\tau) S_{x''+x-x'}^b(0) + ... \\ &= \frac{1}{|\tau|^{2\Delta_S^{0}}} \sum_{x''=1}^L \mathds{1}_{x''} S_{x''+x}^a(\tau) S_{x''+x'}^a(0) + \frac{1}{|\tau|^{2\Delta_S^{0}}} \sum_{x''=1}^L \mathds{1}_{x''} S_{x''+x}^a(\tau) S_{x''-x'}^a(0) \\ &\hspace{0.5cm}+ \frac{1}{|\tau|^{2\Delta_S^{0}}} \sum_{x''=1}^L \mathds{1}_{x''+x} S_{x''}^a(\tau) S_{x''+x+x'}^a(0) + \frac{1}{|\tau|^{2\Delta_S^{0}}} \sum_{x''=1}^L \mathds{1}_{x''+x} S_{x''}^a(\tau) S_{x''+x-x'}^a(0) + ... \\ &\approx  \frac{2}{|\tau|^{2\Delta_S^{0}}} O_{x+x'}(0) +  \frac{2}{|\tau|^{2\Delta_S^{0}}} O_{x-x'}(0) + ... \, ,
	\end{split}
\end{align}

\noindent which leads to the OPE coefficients

\begin{equation}
	C_{xx',x+x'} = C_{xx',x-x'} = C_{xx',x'-x} = 2 \, .
\end{equation}

We must now compute the OPE coefficients involving $O_h$. First

\begin{align}
	\begin{split} \label{Eq:OyOh}
		O_x(\tau) O_h(0) &= \sum_{x',x''=1}^L \sum_{a,b} S_{x'}^a(\tau) S_{x'+x}^a(\tau) S_{x''}^b(0) \\ &= \sum_{x'=1}^L \sum_{a,b} \Big[ S_{x'}^a(\tau) S_{x'}^b(0) \Big] S_{x'+x}^a(\tau) + \sum_{x'=1}^L \sum_{a,b} \Big[ S_{x'+x}^a(\tau) S_{x'+x}^b(0) \Big] S_{x'}^a(\tau) + ... \\ &= \sum_{x'=1}^L \sum_{a,b} \Bigg[ \frac{\delta^{ab}}{|\tau|^{2\Delta_S^{0}}} \mathds{1}_{x'} + i C_{SSS} \sgn(\tau) \frac{f^{abc}}{|\tau|^{\Delta_S^{0}}} S_{x'}^c(0) \Bigg] S_{x'+x}^a(\tau)  \\ &\hspace{0.5cm} + \sum_{x'=1}^L \sum_{a,b} \Bigg[ \frac{\delta^{ab}}{|\tau|^{2\Delta_S^{0}}} \mathds{1}_{x'+x} + i C_{SSS} \sgn(\tau) \frac{f^{abc}}{|\tau|^{\Delta_S^{0}}} S_{x'+x}^c(0) \Bigg] S_{x'}^a(\tau) + ... \\ &= \frac{2 }{|\tau|^{2\Delta_S^{0}}} O_h(0) + ... \, ,
	\end{split}
\end{align}

\noindent where the two terms with $C_{SSS}$ cancel each other. From this, we identify $C_{xhh} = 2$. By proceeding similarly, one can easily show that the $O_h$-$O_h$ OPE does not give anything interesting.

\subsubsection{RG equations}

We are now in a position to compute the RG equations, using Eq.~\ref{Eq:Cardy} and the OPE coefficients previously computed. First, we get for $J_x$

\begin{align}
	\begin{split}
		\beta(J_x) &= -\epsilon J_x - \sum_{x'=1}^{\frac{L-1}{2}} C_{x'x',x} J_{x'}^2 - \sum_{x'\neq x''=1}^{\frac{L-1}{2}} C_{x'x'',x} J_{x'} J_{x''} + ... \, ,
	\end{split}
\end{align}

\noindent where the terms with $x'=x''$ and $x'\neq x''$ have been separated. Using our results from above, the OPE coefficient in the second term is

\begin{align}
	\begin{split}
		C_{x'x',x} &= \delta(x'-x) C_{xx,x} + \Big[ \delta(2x'-x) + \delta(2x'-(L-x)) \Big] C_{xx,2x} \\ &= - N C_{SSS}^2 \delta(x'-x) + 2 \Big[ \delta(2x'-x) + \delta(2x'-(L-x)) \Big] \, ,
	\end{split}
\end{align}

\noindent where the second term contains two $\delta$ functions to account for the fact that $O_{L-x}=O_x$. For the OPE coefficient of the third term, we have

\begin{align}
	\begin{split}
		C_{x'x'',x} &= \Big[ \delta(x-x'-x'')  + \delta(L-x-x'-x'') \Big] C_{x'x'',x'+x''} \\ &\hspace{0.5cm} + \Big[ \delta(x-x'+x'') + \delta(x+x'-x'') \Big] C_{x'x'',x'-x''} \\ &= 2 \Big[ \delta(x-x'-x'') + \delta(L-x-x'-x'') + \delta(x-x'+x'') + \delta(x+x'-x'') \Big] \, ,
	\end{split}
\end{align}

\noindent where this expression accounts for the symmetries $O_{L-x} = O_{x}$ and $O_{-x}=O_x$. Therefore, the RG equation for $J_x$ is

\begin{align}
	\begin{split}
		\beta(J_x) &= -\epsilon J_x+ N C_{SSS}^2 J_x^2 - 2 \sum_{x'=1}^{\frac{L-1}{2}} \Big[ \delta(2x'-x) + \delta(2x'-(L-x)) \Big] J_{x'}^2 \\ &\hspace{0.5cm} -2 \sum_{x'\neq x''=1}^{\frac{L-1}{2}} \Big[ \delta(x-x'-x'') + \delta(L-x-x'-x'') + \delta(x-x'+x'')  + \delta(x+x'-x'') \Big] J_{x'} J_{x''} \, .
	\end{split}
\end{align}

\noindent The last term can be rewritten in a simpler form by restricting the sum in the last term to $x'<x''$, yielding

\begin{align}\label{Eq:beta_J_x}
	\begin{split}
		\beta(J_x) &= -\epsilon J_x + N C_{SSS}^2 J_x^2 - 2 \sum_{x'=1}^{\frac{L-1}{2}} \Big[ \delta(2x'-x) + \delta(2x'-(L-x)) \Big] J_{x'}^2 \\ &\hspace{0.5cm} -4 \sum_{x'=1<x''}^{\frac{L-1}{2}} \Big[ \delta(x-x'-x'') + \delta(L-x-x'-x'') + \delta(x+x'-x'') \Big] J_{x'} J_{x''} \, .
	\end{split}
\end{align}

The RG equation for $h$, following from Eqs.~\ref{Eq:Cardy} and \ref{Eq:OyOh}, is on the other hand

\begin{equation}\label{Eq:beta_h}
	\beta(h) = (1-\Delta_S^{0}) h - 4 \sum_{x'=1}^{\frac{L-1}{2}} J_{x'} h \, .
\end{equation}

\noindent From this, we get the RG eigenvalue $e_h = \frac{\partial \beta(h)}{\partial h}$, which is itself related to the impurity (corrected) scaling dimension $\Delta_S = D-e_h = 1- e_h$, which thus yields

\begin{equation}
	\Delta_S = \Delta_S^{0} + 4 \sum_{x'=1}^{\frac{L-1}{2}} J^{*}_{x'} = \frac{1}{1+\gamma} + 4 \sum_{x'=1}^{\frac{L-1}{2}} J^{*}_{x'} \label{eq:scaling_dim_change_odd}
\end{equation}
\noindent where $J^{*}_{x}$ denotes the fixed-point value of the couplings.

\subsubsection{RG equations for $L$ even}

Let us now consider the case where $L$ is even. We still work with the perturbing operators defined in Eq.~\ref{Eq:O_x}, where this time $x=[1,\frac{L}{2}-1]$. However, a subtlety arises due to the operator $O_{L/2}$, which must be treated separately, since it maps to itself under the transformation $x \rightarrow L-x$. Repeating analogous steps as for $L$ odd, one can show that the RG equations are

\begin{align}\label{Eq:beta_Jx_even}
	\begin{split}
		\beta(J_x) &= -\epsilon J_x + N C_{SSS}^2 J_x^2 - 2 \sum_{x'=1}^{\frac{L}{2}-1} \Big[ \delta(2x'-x)+\delta(2x'+x-L) \Big] J_{x'}^2 \\ &\hspace{0.5cm} - 4 \sum_{x'=1<x''}^{\frac{L}{2}-1} \Big[ \delta(x-x'-x'')+\delta(L-x-x'-x'') + \delta(x+x'-x'') \Big] J_{x'} J_{x''} \\ &\hspace{0.5cm} - 8  \sum_{x'=1}^{\frac{L}{2}-1} \delta(L/2-x'-x) J_{x'} J_{L/2} \, ,
	\end{split}
\end{align}

\begin{align}\label{Eq:beta_JL/2_even}
	\begin{split}
		\beta(J_{L/2}) = -\epsilon J_{L/2} - 2 \sum_{x'=1}^{\frac{L}{2}-1} \delta(x'-L/4) J_{x'}^2 - 4 \sum_{x'=1<x''}^{\frac{L}{2}-1} \delta(L/2-x'-x'') J_{x'} J_{x''} + 2 N C_{SSS}^2 J_{L/2}^2 \, ,
	\end{split}
\end{align}

\begin{equation}
	\beta(h) = (1-\Delta_S^0) h - 4 \sum_{x'=1}^{\frac{L}{2}-1} J_{x'} h - 4 J_{L/2} h \, .
\end{equation}

\noindent The scaling dimension of the impurity is thus

\begin{equation}
	\Delta_S = \Delta_S^{0} + 4 \sum_{x'=1}^{\frac{L}{2}-1} J_{x'}^* + 4 J_{L/2}^* = \frac{1}{1+\gamma} + 4 \sum_{x'=1}^{\frac{L}{2}-1} J_{x'}^* + 4 J_{L/2}^* \, . 
	\label{eq:scaling_dim_change_even}
\end{equation}

\subsubsection{OPE coefficient $C_{SSS}$ at large $N$ and $K$ in the bCFT}

The OPE coefficient $C_{SSS}$ can easily be computed in the limit $N,K \rightarrow \infty$, with $\frac{K}{N}=\gamma$ fixed. To do so, the impurity is written as $S^a(\tau) = f_{\alpha}^{\dag}(\tau) T_{\alpha\beta}^a f_{\beta}(\tau)$ and we use the result of Ref. \cite{parcollet1998overscreened} for the spinon imaginary-time Green's function $G_f(\tau)$, defined as

\begin{equation}
	\delta_{\alpha \beta} G_f(\tau) = -\langle T_{\tau}\hat{f}_{\alpha}(\tau) \hat{f}_{\beta}^{\dag}(0) \rangle = \delta_{\alpha \beta} \sgn(\tau) \frac{A_f}{|\tau|^{2\Delta_f}} \, .
\end{equation}

\noindent The real coefficient $A_f$ is obtained by computing the impurity 2-point function and demanding that $C_{SS\mathds{1}} = 1$. In the large $N,K$ limit, the action is quadratic in the spinons $f_{\alpha}$ \cite{parcollet1998overscreened} and Wick's theorem can thus be used. Therefore

\begin{align}
	\begin{split}
		\langle S^a(\tau) S^b(0) \rangle = - \Tr[T^a T^b] G_f(-\tau) G_f(\tau) =  \frac{A_f^2}{2} \frac{\delta^{ab}}{|\tau|^{2\Delta_S^0}} \, ,
	\end{split}
\end{align}

\noindent where we have used the generator normalization condition $\Tr[T^a T^b] = \frac{1}{2} \delta^{ab}$. Hence, it follows that $A_f = \sqrt{2}$. The OPE coefficient $C_{SSS}$ is now obtained by multiplying Eq.~\ref{Eq:OPE_BCFT} by $S^c(\tau')$ on the right side and by taking the expectation value in the bCFT, resulting in

\begin{equation} \label{Eq:SSS}
	\langle S^a(\tau) S^b(0) S^c(\tau') \rangle = i C_{SSS} \sgn(\tau) \frac{f^{abc}}{|\tau|^{\Delta_S^0} |\tau'|^{2\Delta_S^0}} \, ,
\end{equation}

\noindent where $|\tau'| \gg |\tau|$. Computing the impurity 3-point function yields two Wick contractions and results in

\begin{align}
	\begin{split}
		\langle S^a(\tau) S^b(0) S^c(\tau') \rangle &= \Tr[T^b T^a T^c] G_f(-\tau) G_f(\tau-\tau') G_f(\tau') + \Tr[T^a T^b T^c] G_f(\tau) G_f(\tau'-\tau) G_f(-\tau') \\ &= \Tr\Big( [T^a,T^b] T^c \Big) G_f(\tau) G_f(\tau') G_f(\tau-\tau') \\ &= \frac{i}{2} f^{abc} 2^{3/2} \frac{\sgn(\tau) \sgn(\tau') \sgn(\tau-\tau')}{|\tau|^{\Delta_S^0} |\tau'|^{\Delta_S^0} |\tau-\tau'|^{\Delta_S^0}} \\ &\approx -\sqrt{2} i f^{abc} \frac{\sgn(\tau)}{|\tau|^{\Delta_S^0} |\tau'|^{2\Delta_S^0}} \, ,
	\end{split}
\end{align}

\noindent where we have used the antisymmetry of $G_f$ under $\tau\rightarrow -\tau$, the SU$(N)$ algebra $[T^a,T^b]=i f^{abc} T^c$, the generator normalization condition and the fact that $|\tau'|\gg |\tau|$. Hence, by comparing with the right side of Eq.~\ref{Eq:SSS}, we see that 

\begin{equation}
	C_{SSS} = -\sqrt{2} + \mathcal{O}(1/N) \, .
\end{equation}

\subsection{Solutions and universal quantities}

\begin{figure}[h]
	\begin{center}
		\includegraphics[width=0.8\columnwidth,angle=-0]{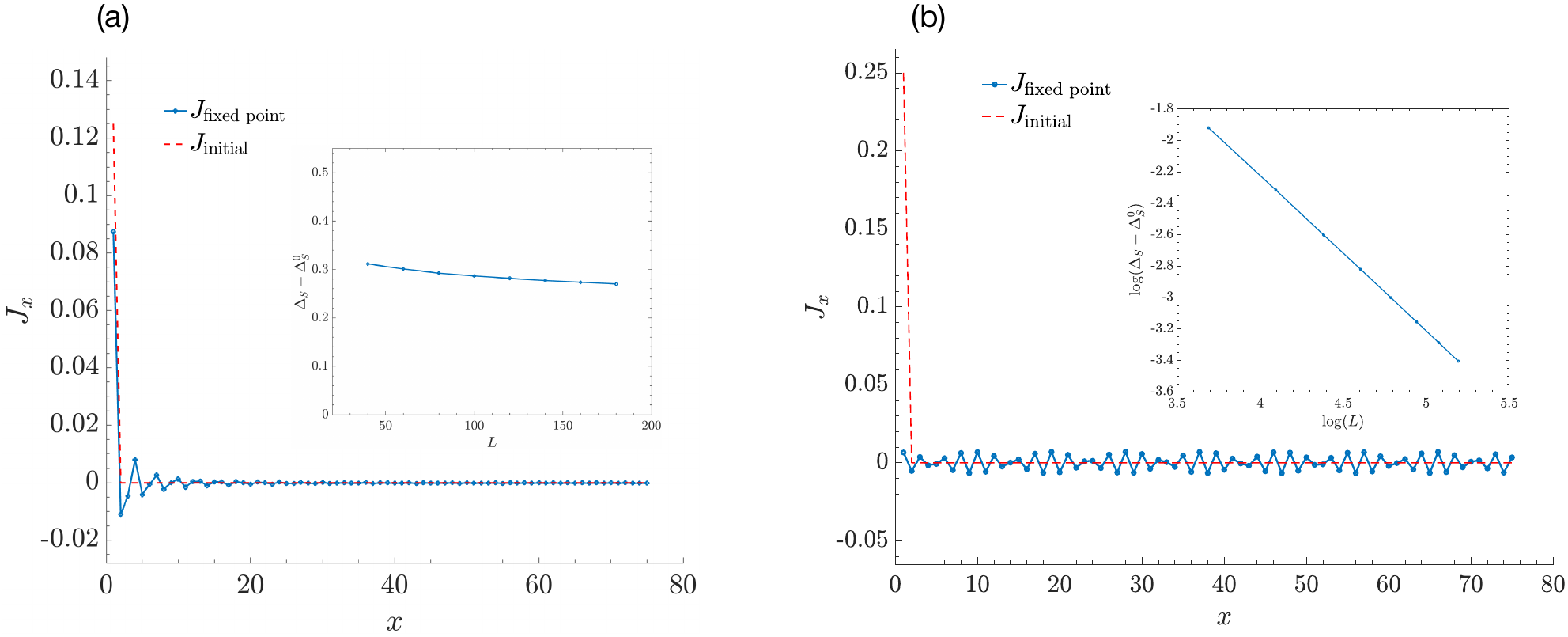}
		\caption{Two characteristic fixed points of RG with one relevant direction. (a) When $\alpha = N C^2_{SSS} \gtrsim 8$, $|J_x|$ decays exponentially with oscillating sign structure. The figure shows the fixed point values of $J_x$ as a function of $x$ for $\alpha = 8, \epsilon = 1$ at a system size $L = 60$. The inset shows the 1-loop correction to the scaling dimension of the impurity,  associated with the temporal spin-spin correlations, at the critical point for different system sizes. (b) When $\alpha = \mathcal{O}(1)$, one finds a different fixed point with a single relevant direction where $|J_x| \sim 1/L$ with an oscillating sign structure. The figure shows the solution for $\alpha = 4, \epsilon = 1$ at a system size $L = 60$.  Both of these solutions are reachable by starting with short-range interactions where $J_{x=1}$ is the only non-zero coupling (as depicted by the dashed curve $J_{\textrm{initial}})$.}
		\label{fig:RG_solns}
	\end{center}
\end{figure}

The goal is now to solve the RG equations (Eq. \ref{Eq:beta_J_x} for $L$ odd or Eqs. \ref{Eq:beta_Jx_even} and \ref{Eq:beta_JL/2_even} for $L$ even) to identify the fixed point with a single relevant direction describing the transition between the decoupled and the dimer phases. Without any constraints, the number of fixed points grows rapidly with $L$. Since our microscopic Hamiltonian is local, we restrict ourselves to solutions that can be obtained by starting with an initial value of $\{J_x\}$ that  describes a local Hamiltonian (e.g. $J_{x\neq 1} = 0$ describes one such initial condition). We search for such solutions numerically by evolving the $\beta$-functions along the RG flow, and further restricting attention to solutions that have a single relevant direction. Fig.~\ref{fig:RG_solns} shows two different classes of solutions that generically appear over a range of parameters. One such solution (Fig.~\ref{fig:RG_solns} (a)) appears when the parameter $\alpha = N C^2_{SSS} \gtrsim 8$ and corresponds to an exponentially decaying $J_x$ with oscillating signs. The other solution (Fig.~\ref{fig:RG_solns} (b)) appears when $\alpha$ is small, and corresponds to a non-decaying $J_x$ with oscillating signs. However, for this latter set of solutions, the maximum value of $|J_x|$ decays with the total system size approximately as $1/L$, and therefore, in a strict sense, all $J_x$ vanish in the thermodynamic limit. The change in the temporal scaling dimension, proportional to $\sum_x J_x^*$ at the fixed point (Eq.~\ref{eq:scaling_dim_change_even}) also appears to decay as $1/L$.

\section{Details of the Monte Carlo simulations}\label{Sec:Appendix_B}

\subsection{Formulation and Wilson discretization of the bath}
We consider a lattice of  SU($N$) $K$-channel Kondo models in the  totally analytically-symmetric representation corresponding to a Young tableau with  $N/2$ lines and a single column. 
To each lattice site, $x$, we attach  $K$  SU($N$) channels   that  will  screen an SU($N$) spin  degree of freedom.   For the  fermionic  baths we introduce the creation operator 
$\hat{c}^{\dagger}_{i\alpha}(x,k)$  where $i = 1 \cdots K$ is the channel index, $\alpha = 1 \cdots N$ is the spin index, $k = 1 \cdots  L_b$ is  a one-dimensional momentum, and $x = 1 \cdots L$ labels the unit cell.   Our aim is to formulate an 
efficient auxiliary field quantum Monte Carlo formulation of the Hamiltonian: 
\begin{equation}
	\hat{H}=  \hat{H}_0  + \hat{H}_{J_K}  + \hat{H}_{J_H}
\end{equation} 
with 
\begin{eqnarray}
	\hat{H}_0  &= &\sum_{i=1}^{K}\sum_{\alpha=1}^{N} \sum_{k=1}^{L_b} \sum_{x=1}^{L} \epsilon(k) \hat{c}^{\dagger}_{i\alpha}(x,k) \hat{c}^{}_{i\alpha}(x,k)   \nonumber \\ 
	\hat{H}_{J_K} 	& = &  \frac{2J_K}{N} \sum_{x=1}^{L} \sum_{a=1}^{N^2-1} \sum_{\alpha,\alpha'=1}^{N} \sum_{i=1}^{K}\hat{c}^{\dagger}_{i\alpha}(x,y=0) T^a_{\alpha\beta}\hat{c}^{}_{i\beta}(x,y=0) \,\, \hat{S}^a(x) \, \, \,   \text{and} \nonumber \\
	\hat{H}_{J_H}  & =  &  \frac{2J_H}{N} \sum_{x=1}^{L} \sum_{a=1}^{N^2-1}  \hat{S}^a(x) \hat{S}^a(x+1). 
\end{eqnarray} 
In the above,  $T^{a}$ are generators  of  SU($N$) satisfying the normalization condition: 
\begin{equation}
	\Tr(T^a T^b) = \frac{1}{2} \delta_{ab}.
\end{equation}

To formulate the quantum Monte Carlo algorithm, we will adopt a fermionic representation of the SU($N$) spin degree of freedom, 
\begin{equation}
	\hat{S}^a(x) = \sum_{\alpha,\beta=1}^{N} \hat{f}^{\dagger}_{\alpha}(x) T^a_{\alpha\beta} \hat{f}^{}_{\beta}(x)
\end{equation}  
with the constraint: 
\begin{equation}
	\sum_{\alpha=1}^{N}\hat{f}^{\dagger}_{\alpha}(x) \hat{f}^{}_{\alpha}(x) = \frac{N}{2}.
\end{equation}
We impose the  constraint energetically  thereby adding a Hubbard-$U$ term to the Hamiltonian:
\begin{equation}
	\hat{H}_{QMC}=  \hat{H}_0  + \hat{H}_{J_K}  + \hat{H}_{J_H}  + \hat{H}_U
\end{equation}
such that in the large-$U$ limit $\left. \hat{H}_{QMC} \right|_{U = \infty}=  \hat{H}$. 

We proceed by describing how we implement each term in the  ALF-formulation of the  projective auxiliary field 
quantum Monte Carlo algorithm. 
For the  Wilson discretization of the bath, we rewrite the kinetic energy term as: 
\begin{eqnarray}
	\hat{H}_0  &= &\sum_{i=1}^{K}\sum_{\alpha=1}^{N} \sum_{k=1}^{L_b} \sum_{x=1}^{L} \epsilon(k) \hat{c}^{\dagger}_{i\alpha}(x,k) \hat{c}^{}_{i\alpha}(x,k)   \nonumber \\ 
	& = & \sum_{i=1}^{K}\sum_{\alpha=1}^{N} \sum_{k>0}  \sum_{x=1}^{L} \epsilon(k) \left( \hat{c}^{\dagger}_{i\alpha}(x,k)  \hat{c}^{}_{i\alpha} (x, k)  
	+  \hat{c}^{\dagger}_{i\alpha} (x,-k) \hat{c}^{}_{i\alpha} (x,-k)  \right)  \nonumber \\ 
	& = &  \sum_{i=1}^{K}\sum_{\alpha=1}^{N} \sum_{k>0} \sum_{x=1}^{L}  \epsilon(k) \left( \hat{\gamma}^{\dagger}_{i\alpha}(x,k) \hat{\gamma}^{}_{i\alpha}(x,k)  
	+  \hat{\eta}^{\dagger}_{i\alpha}(x,k) \hat{\eta}^{}_{i\alpha}(x,k)  \right) 
\end{eqnarray}
with 
\begin{eqnarray}
	& & \hat{\gamma}^{\dagger}_{i\alpha}(x,k)   = \frac{1}{\sqrt{2}} \left(  \hat{c}^{\dagger}_{i\alpha}(x,k) + \hat{c}^{\dagger}_{i\alpha}(x,-k)\right)  \nonumber \\
	& & \hat{\eta}^{\dagger}_{i\alpha}(x,k)   = \frac{1}{\sqrt{2}} \left(  \hat{c}^{\dagger}_{i\alpha}(x,k) - \hat{c}^{\dagger}_{i\alpha}(x,-k)\right).
\end{eqnarray}
In the  above,   we  have assumed that $\epsilon(k) = \epsilon(-k)$. 

We can now replace the sum over momenta by: 
\begin{equation}
	\frac{L_b}{2} \equiv \sum_{k>0} =  \int d \epsilon g(\epsilon)  \text{   with  }  g(\epsilon) =  \sum_{k > 0 }  \delta( \epsilon - \epsilon(k) )  
\end{equation}
and choose: 
\begin{equation}
	g(\epsilon)  = \left\{
	\begin{array}{ll}
		\frac{L_b}{2 W}    &     \text{ if } | \epsilon | < W/2 \\
		0   & \text{ otherwise. }
	\end{array}
	\right.
\end{equation}
Choosing an arbitrary discretization,  $   \int d \epsilon g(\epsilon)  f(\epsilon) =  
\sum_{n=1}^{L_b}  \Delta \epsilon_n  g(\epsilon_n )  f(\epsilon_n) $,  the non-interacting part  reads: 
\begin{equation}
	\hat{H}_0   =  \sum_{i=1}^{K}\sum_{\alpha=1}^{N}  \sum_{n=1}^{L_b}  \sum_{x=1}^{L} \Delta \epsilon_n  \epsilon_n g(\epsilon_n )  \left( \hat{\gamma}^{\dagger}_{i\alpha} (x,n) \hat{\gamma}^{}_{i\alpha} (x,n)  
	+  \hat{\eta}^{\dagger}_{i\alpha}(x,n) \hat{\eta}^{}_{i\alpha}(x,n)  \right). 
\end{equation}
Here $L_b$ corresponds to  the number of energy values $\epsilon_n$. As mentioned previously, the energy  discretization of the bath is arbitrary. 
Here, we have opted  for a  logarithmic  Wilson-like discretization defined by: 
\begin{equation}
	\label{eq.Wilson}
	\epsilon_n = \begin{cases}
		-\frac{W}{2}\Lambda^{1-n}  & \text{ for } n = 1 \cdots L_b/2 \\
		+\frac{W}{2} \Lambda^{n-{L_b} } & \text{ for } n = L_b/2+1 \cdots L_b.
	\end{cases}
\end{equation}
In the simulations we have  typically used $\Lambda = 1.5$ and $L_b=30$ chosen by benchmarking our data against  the single impurity case, see Sec.~\ref{sec:Bench}.

We now  discuss the Kondo term.   Consider the hybridization: 
\begin{equation}
	\label{Hyb.eq}
	\hat{V}^{\dagger}_{i}(x)    =  \sum_{\alpha} \hat{f}^{\dagger}_{\alpha}(x)
	\underbrace{\frac{1}{\sqrt{L_b}}\sum_{k=1}^{L_b} 
		\hat{c}^{\phantom\dagger}_{i\alpha}(x,k)}_{\hat{c}^{}_{i\alpha}(x,y=0)}.
\end{equation}
The  Kondo interaction then reads: 
\begin{equation}
	\hat{H}_{J_K} =  -\frac{J_K}{2 N}  \sum_{i=1 }^{K} \sum_{x=1}^{L}  \left(
	\hat{V}^{\dagger}_{ i}(x) \hat{V}^{\phantom\dagger}_{i}(x)   +
	\hat{V}^{\phantom\dagger}_{ i}(x) \hat{V}^{\dagger}_{ i}(x)  \right)  \equiv 
	-\frac{J_K}{4 N}  \sum_{i=1 }^{K} \sum_{x=1}^{L}   \left(
	( \hat{V}^{\dagger}_{ i}(x) +  \hat{V}^{\phantom\dagger}_{i}(x) )^2    + ( i\hat{V}^{\dagger}_{ i}(x) - i \hat{V}^{\phantom\dagger}_{i}(x) )^2
	\right)   
\end{equation}   
with 
\begin{eqnarray}
	\hat{V}^{\dagger}_{ i}(x) +  \hat{V}^{\phantom\dagger}_{i}(x)&= &  
	\frac{1}{ \sqrt{L_b} }\sum_{\alpha=1}^{N} \sum_{k=1}^{L_b} \left( \hat{c}^{\dagger}_{i\alpha}(x,k) \hat{f}^{}_{\alpha}(x) +  \hat{f}^{\dagger}_{\alpha}(x) \hat{c}^{}_{i\alpha}(x,k)  \right)  \nonumber \\ & = &
	\frac{1}{ \sqrt{L_b} }\sum_{\alpha=1}^{N} \sum_{k>0} \left( \left[ \hat{c}^{\dagger}_{i\alpha}(x,k) + \hat{c}^{\dagger}_{i\alpha}(x,-k) \right] \hat{f}^{}_{\alpha}(x) +  \hat{f}^{\dagger}_{\alpha}(x) \left[ \hat{c}^{}_{i\alpha}(x,k) +  \hat{c}^{}_{i\alpha}(x,-k) \right] \right) \nonumber \\
	& = &  \frac{\sqrt{2}}{ \sqrt{L_b} }\sum_{\alpha=1}^{N} \sum_{k>0} \left( \hat{\gamma}^{\dagger}_{i\alpha}(x,k) \hat{f}^{}_{\alpha}(x) +  \hat{f}^{\dagger}_{\alpha}(x) \hat{\gamma}^{}_{i\alpha}(x,k) \right)  \nonumber \\ 
	&=& \sqrt{ \frac{2}{L_b} } \sum_{\alpha=1}^{N} \sum_{n=1}^{L_b}  
	\Delta \epsilon_n g(\epsilon_n) \left( \hat{\gamma}^{\dagger}_{i\alpha}(x,n) \hat{f}^{}_{\alpha}(x) +  \hat{f}^{\dagger}_{\alpha}(x) \hat{\gamma}^{}_{i\alpha}(x,n) \right) 
\end{eqnarray}
and 
\begin{eqnarray}
	i\hat{V}^{\dagger}_{ i}(x)  -i  \hat{V}^{\phantom\dagger}_{i}(x)&= &  \sqrt{ \frac{2}{L_b} } \sum_{\alpha=1}^{N} \sum_{n=1}^{L_b}  \Delta \epsilon_n g(\epsilon_n) \left( i \hat{\gamma}^{\dagger}_{i\alpha}(x,n) \hat{f}^{}_{\alpha}(x) -i \hat{f}^{\dagger}_{\alpha}(x) \hat{\gamma}^{}_{i\alpha}(x,n) \right).
\end{eqnarray}
As one can see, $\hat{\eta}^{\dagger}_{i\alpha}(x,k)$  does not couple to the  spin degree of freedom. 
To impose the constraint
\begin{equation}
	\sum_{\alpha=1}^{N}  \hat{f}^{\dagger}_{\alpha}(x)  \hat{f}^{}_{\alpha}(x)  = \frac{N}{2},
\end{equation}
we include  a  Hubbard interaction
\begin{equation}
	\hat{H}_U = \frac{U}{N} \sum_{x=1}^{L}\left( \sum_{\alpha=1}^{N}  \left(  \hat{f}^{\dagger}_{\alpha}(x)  \hat{f}^{}_{\alpha}(x)  - \frac{1}{2}  \right)  \right)^2.
\end{equation}
Importantly  
\begin{equation}
	\left[ \hat{H}_U, \hat{H} \right]  =  0  
\end{equation}
such that the constraint is imposed very efficiently. 
The last element of the Hamiltonian is the Heisenberg interaction between the  SU($N$)  spin  1/2  degrees of freedom:
\begin{equation}
	\hat{H}_{J_H} =  -\frac{J_H}{2 N}  \sum_{x=1}^{L}  \left(
	\hat{D}^{\dagger}_{ x, x+1} \hat{D}^{\phantom\dagger}_{x,x+1}   +
	\hat{D}^{\phantom\dagger}_{ x,x+1 } \hat{D}^{\dagger}_{ x,x+1}  \right)  \equiv 
	-\frac{J_H}{4 N}  \sum_{x=1}^{L}   \left(
	( \hat{D}^{\dagger}_{ x,x+1 } +  \hat{D}^{\phantom\dagger}_{x,x+1} )^2    + 
	( i\hat{D}^{\dagger}_{ x,x+1} - i \hat{D}^{\phantom\dagger}_{x,x+1} )^2 
	\right)   	
\end{equation}
where 
\begin{equation}
	\hat{D}^{\dagger}_{ x, x+1} = \sum_{\alpha=1}^{N} \hat{f}^{\dagger}_{\alpha}(x) \hat{f}^{}_{\alpha}(x+1).
\end{equation}

All interaction terms are  in the form of a perfect square such  that the model can readily be implemented using the Algorithm for Lattice Fermion (ALF) \cite{ALF_v2} library.   Here we  use the Hubbard-Stratonovitch transformation 
\begin{equation}
	e^{\Delta \tau \alpha \hat{A}^2}  \propto  \int d \Phi e^{- \Delta \tau \left( \frac{\Phi^2}{4 | \alpha|} +  \Phi \sqrt{\text{sign}(\alpha)} \hat{A} \right) }
\end{equation}
for  $\alpha \in \mathbb{R}$ and $\Delta \tau$  an imaginary time time step.  Note that in the ALF implementation we use the  Gauss–Hermite quadrature  to replace the continuous variable by a discrete one.  The imaginary time propagator then reads:
\begin{eqnarray}
	e^{- \beta \hat{H}} && \propto  \int D \left\{ \chi_{x,x+1}(\tau)\right\} D \left\{ V_{i,x}(\tau)\right\}   
	D \left\{ \Phi_{x}(\tau)\right\} \\
	&& e^{-N\int_{0}^{\beta} d \tau \left( \sum_{x}\frac{|\chi_{x,x+1}(\tau)|^2}{J_H} +\sum_{x,i} \frac{|V_{i,x}(\tau)|^2}{J_K} + \sum_{x}\frac{|\Phi_{x}(\tau)|^2}{4U} \right)}  \prod_{\alpha=1}^{N} {\cal T} e^{- \int_0^{\beta} d \tau  \hat{h}_{\alpha}(\chi,V,\Phi) }. \nonumber
\end{eqnarray}
In the  above  we have introduced a complex bond fields, $\chi_{x,x+1}(\tau)$ and $V_{i,x}(\tau)$  to decouple the Heisenberg term and Kondo terms respectively and the Hubbard interaction is decoupled using a real field $\Phi_{x}(\tau)$. $ {\cal T} $ denotes the time ordering operator, and the single particle  Hamiltonian reads:
\begin{eqnarray}
	\label{eq.single}
	\hat{h}_{\alpha}(\chi,V,\Phi) =  & & \sum_{i=1}^{K} \sum_{n=1}^{L_b} \sum_{x=1}^{L}  \Delta \epsilon_n \epsilon_n g(\epsilon_n)  \hat{\gamma}^{\dagger}_{i\alpha}(x,n) \hat{\gamma}^{}_{i\alpha}(x,n)  
	+  \sum_{x} \left( \chi_{x,x+1}(\tau) \hat{f}^{\dagger}_{\alpha}(x) \hat{f}^{}_{\alpha}(x+1)  + \text{H.c.} \right)   \\
	&&  \sum_{i=1}^{K} \sum_{x=1}^{L} \left( V_{i,x}(\tau)
	\sqrt{ \frac{2}{L_b} } \sum_{n=1}^{L_b}  \Delta \epsilon_n g(\epsilon_n) \hat{\gamma}^{\dagger}_{i\alpha}(x,n) \hat{f}^{}_{\alpha}(x)    + \text{H.c.}  \right) + \sum_{x=1}^{L} i \Phi_{x}(\tau)  \left(  \hat{f}^{\dagger}_{\alpha}(x)  \hat{f}^{}_{\alpha}(x)  - \frac{1}{2}  \right). \nonumber
\end{eqnarray}
We have used the projective formulation of the auxiliary field quantum Monte Carlo algorithm \cite{Sorella89,Sugiyama86,AssaadEvertz2008}: 
\begin{equation}
	\frac{\langle \Psi_0 | \hat{O} | \Psi_0 \rangle}{\langle \Psi_0 | \Psi_0 \rangle} = \lim_{\theta \rightarrow \infty} \frac{\langle \Psi_T | e^{-\theta \hat{H}} \hat{O} e^{-\theta \hat{H}} | \Psi_T \rangle}{\langle \Psi_T | e^{-2 \theta \hat{H}} | \Psi_T \rangle}.
\end{equation} 
The above holds under the condition that the trial wave function $| \Psi_T \rangle $ is not orthogonal to the ground state. The trial wave function is chosen to be a single Slater determinant and  to respect symmetries of the model. As we will detail below, this includes channel and spin symmetry as well as U(1) charge.  We generate the trial wave function by solving the single particle Hamiltonian of Eq.~\ref{eq.single} for a space-time constant  value of $\chi_{x,x+1}(\tau)$ and vanishing values of $V_{i,x}(\tau)$ and $\Phi_x(\tau)$.    Note that setting $V_{i,x}(\tau)$ to zero stems from the requirement of satisfying channel symmetry. Furthermore, for $\Phi_{x}(\tau) =0$  particle-hole symmetry is present. 
The trial wave function hence takes the form: 
\begin{equation}
	| \Psi_T\rangle =  \otimes_{\alpha=1}^{N}| \Psi_T, \alpha \rangle.
\end{equation}
In the Monte Carlo simulation we sample configurations according to the \textit{weight}:
\begin{equation}
	W(\chi,V,\Phi)
	= e^{-N \left[\int_{0}^{2\theta} d \tau \left( \sum_{x}\frac{|\chi_{x,x+1}(\tau)|^2}{J_H} +\sum_{x,n} \frac{|V_{x,n}(\tau)|^2}{J_K} + \sum_{x}\frac{|\Phi_{x}(\tau)|^2}{4U} \right)  - \log \langle \Psi_T,1 | {\cal T} e^{-\int_{0}^{2\theta} d \tau  \hat{h}_1(\chi,V,\Phi)} | \Psi_T,1 \rangle \right]}.  
\end{equation}
Since the trial wave function as well as the  Hubbard-Stratonovitch transformation do not break the SU($N$) spin symmetry, $N$ becomes a multiplicative  factor in front of the action. 
Using the particle-hole transformation: 
\begin{equation}
	\hat{\gamma}^{\dagger}_{i1}(x,n) \rightarrow (-1)^{x} \hat{\gamma}^{\phantom\dagger}_{i1}( x,\overline{n}) 
\end{equation}
and 
\begin{equation}
	\hat{f}^{\dagger}_{1}(x) \rightarrow (-1)^{x} \hat{f}^{\phantom\dagger}_{1}(x) 
\end{equation}
with $\epsilon_{\overline{n}} = - \epsilon_n$  and $\Delta \epsilon_{\overline{n}} = \Delta \epsilon_{n}$, we can show that 
$ \langle \Psi_T,1 | {\cal T} e^{-\int_{0}^{2\theta} d \tau  \hat{h}_1(\chi,V,\Phi)} | \Psi_T,1 \rangle $ is real. Hence,  
$W(\chi,V,\Phi)$ is  positive for even values of $N$.

\subsection{\label{sec:Bench} Benchmarking for the single impurity case}
\begin{figure}[h]
	\begin{center}
		\includegraphics[width=0.4\columnwidth,angle=-0]{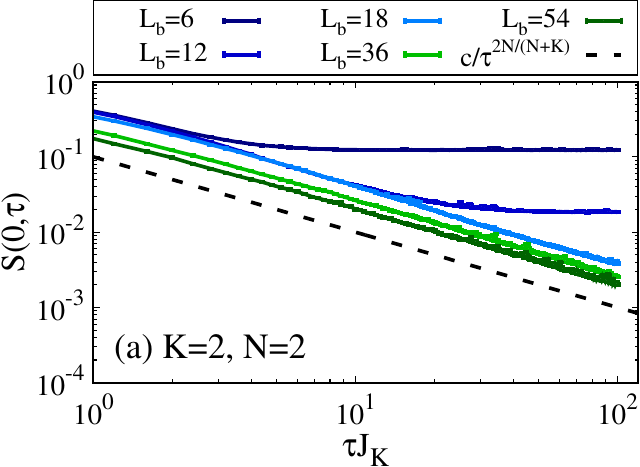}
		\includegraphics[width=0.4\columnwidth,angle=-0]{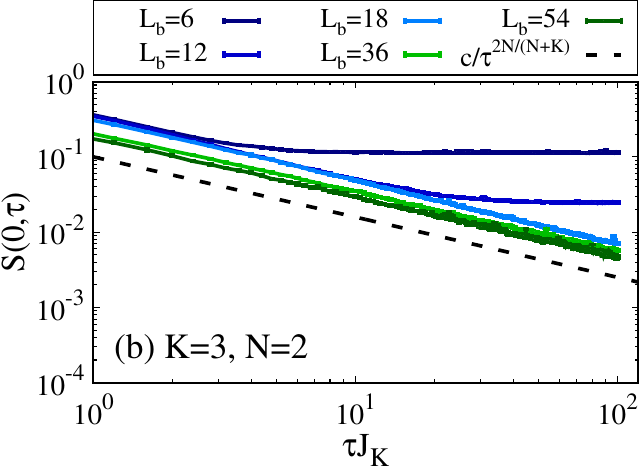}\\
		\includegraphics[width=0.4\columnwidth,angle=-0]{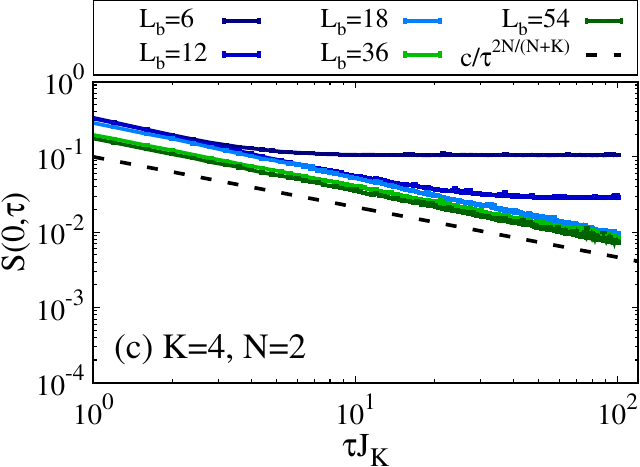}
		\includegraphics[width=0.4\columnwidth,angle=-0]{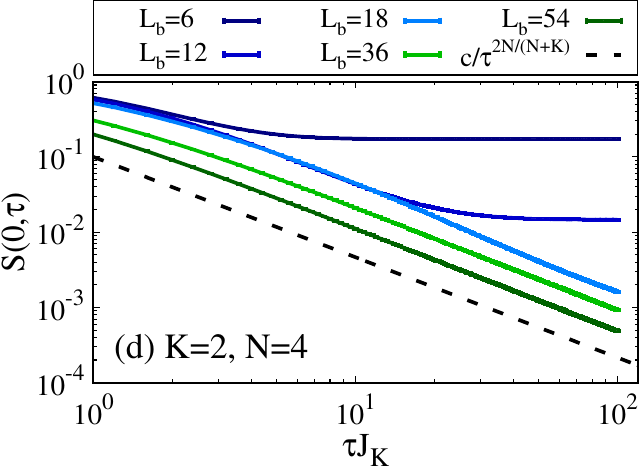}
		\caption{Local spin-spin correlation function $S(0,\tau)$ for the  $K=2$, $K=3$, and $K=4$ channel 
			Kondo impurity models with $N$ spin flavors. Here we consider  a flat  conduction electron density of states 
			with width $W=1$, Kondo coupling $J_K/W=1$, and $L_b$ bath sites per channel.    
			We use a  logarithmic Wilson-type  discretization and the QMC projective parameter $\Theta J_K=100$.   
			The dashed line corresponds to  the form $ c/\tau^{\frac{2N}{N+K}}$  and shows 
			that for $L_b \geq 18$   the data follows the expected behavior  \cite{parcollet1998overscreened}. 
		}
		\label{fig:KN_CK}
	\end{center}
\end{figure}

Figure~\ref{fig:KN_CK} shows that with a minimal number of bath sites $L_b$, we can reproduce the results  of Ref.~\cite{parcollet1998overscreened} for the asymptotic behavior of the  local  
spin-spin correlation function for various values of $N$ and $K$.  We note that these results are prohibitively  hard to obtain  had we chosen a linear discretization of the density of states,  or a  real-space tight binding formulation for  the bath. 

\subsection{Symmetries and observables}
\subsubsection{Channel Symmetry}
\begin{figure}[h]
	\begin{center}
		\includegraphics[width=0.45\columnwidth,angle=-0]{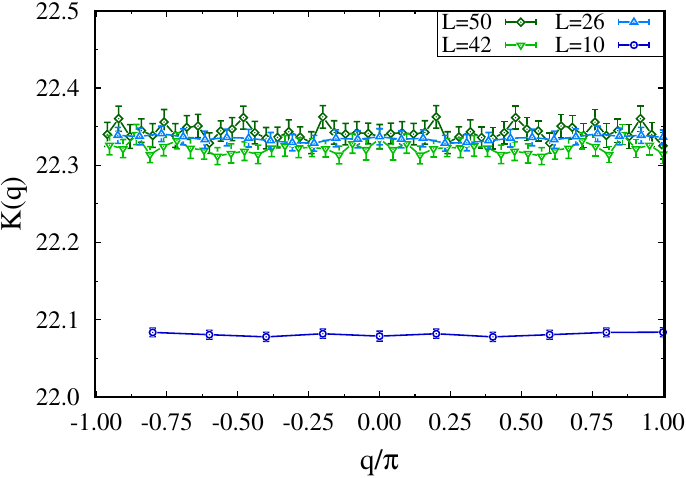}
		\caption{$K(q)$ as defined in Eq.~\ref{eq.Kq} as a function of the Heisenberg chain length $L$ in the VBS 
			phase at $J_H/J_K=0.8$. 
			In the QMC simulations we use $L_b=30$ bath sites per channel, Trotter time step $\Delta\tau J_K=0.1$, and 
			the $L$ dependent projector parameter $\Theta J_K=100$ ($L=10$), $\Theta J_K=150$ ($L=26$), 
			and $\Theta J_K=200$ for the longest chains. 
		}
		\label{fig:channel_symm}
	\end{center}
\end{figure}

Let $T^{m}$  be  the $m:1\cdots K^2 -1 $  generators   of  SU($K$)  with normalization
\begin{equation}
	\label{eq.norm}
	\text{Tr} \, T^{m} T^{m'} =  \frac{1}{2} \delta_{m,m'}
\end{equation}
and 
\begin{equation}
	\hat{T}(\ve{e},\theta)  = \exp\left[ i \sum_{i,i',\alpha,k,x}\hat{\gamma}^{\dagger}_{i\alpha}(x,k) \left(\theta  \ve{e} \cdot \ve{T} \right)_{ii'} \hat{\gamma}_{i'\alpha}(x,k) \right]. 
\end{equation}
Thereby,  
\begin{equation}
	\hat{T}^{\dagger}(\ve{e},\theta) 	\hat{\gamma}^{\dagger}_{i\alpha}(x,k)  \hat{T}^{\phantom\dagger}(\ve{e},\theta)     =  \sum_{i'} \left[ e^{i \theta\ve{e} \cdot \ve{T}  }\right]_{ii'}\hat{\gamma}^{\dagger}_{i'\alpha}(x,k).  
\end{equation}
Clearly,   $ \left[ \hat{H},\hat{T}(\ve{e},\theta) \right]   = 0  $ such  that the Hamiltonian is invariant under an SU($K$) channel rotations. 
Since the $f$-fermion does not carry  a channel index, the  hybridization $\hat{V}_{i,x}$  transforms as $\hat{\gamma}^{\dagger}_{i\alpha}(x,k) $ under an SU($K$)  channel  rotation.  To probe for channel symmetry breaking, we consider:
\begin{equation}
	\hat{O}^{m}(x) = \sum_{i,i'}\hat{V}^{\dagger}_{i}(x) T^{m}_{ii'} \hat{V}^{\phantom\dagger}_{i'}(x). 
\end{equation}
Under a channel  rotation $\hat{O}^{m}(x)  \rightarrow \sum_{m'} R_{mm'}  \hat{O}^{m'}(x)  $  with 
R an SO($K^2 -1$) matrix.  Hence, a non-vanishing  expectation value $ \langle \hat{O}^{m}(x) \rangle $ implies 
channel  symmetry  breaking.  In the code,  we  compute: 
\begin{equation}
	\label{eq.Kq}
	K(q)  =  \sum_{x} \sum_{m}  e^{iqx} \langle  \hat{O}^{m}(x)   \hat{O}^{m}(0)  \rangle  
\end{equation}
and  check for long ranged  order. Note  that the background vanishes by symmetry.  Fig.~\ref{fig:channel_symm}   plots $K(q)$ in the VBS phase at $J_H/J_K = 0.8$ as a function of  $L$. As apparent 
the results are consistent with vanishing real  space correlations. Hence, the channel symmetry is not broken.

\subsubsection{SU($N$) spin symmetry}
The generator of  global SU($N$)  spin rotations  is  noting but the total spin: 
\begin{equation}
	\hat{S}^{a}  = \sum_{i,\alpha,\alpha',k,x}\hat{\gamma}^{\dagger}_{i\alpha}(x,k) T^{a}_{\alpha\alpha'} 
	\hat{\gamma}^{\phantom\dagger}_{i\alpha'}(x,k)  + \sum_{\alpha, \alpha',x} \hat{f}^{\dagger}_{\alpha}(x) T^{a}_{\alpha\alpha'} \hat{f}^{\phantom\dagger}_{\alpha'}(x).  
\end{equation}
In the above $T^{a}$  are the generators of  SU($N$)  with normalization   following that of Eq.~\ref{eq.norm}.   
One will  readily  check that the Hamiltonian commutes with  each component of the spin operator.   To check for spin symmetry breaking, we consider  spin-spin correlation  functions: 
\begin{equation}
	S(x) = \sum_{a} \langle  \hat{S}^{a}_{f}(x)  \hat{S}^{a}_{f}(x=0) \rangle 
\end{equation}
with $\hat{S}^{a}_{f}(x)=\sum_{\alpha, \alpha',x} \hat{f}^{\dagger}_{\alpha}(x) T^{a}_{\alpha\alpha'} \hat{f}^{\phantom\dagger}_{\alpha'}(x)$. 

To detect translation symmetry breaking, we have computed dimer-dimer correlation functions  given by:

\begin{equation}
	D(x-x')= \langle   \ve{\hat{S}}_{f,x} \cdot \ve{\hat{S}}_{f,x+1}   \ve{\hat{S}}_{f,x'} \cdot \ve{\hat{S}}_{f,x'+1} \rangle  - 
	\langle   \ve{\hat{S}}_{f,x} \cdot \ve{\hat{S}}_{f,x+1} \rangle \langle   \ve{\hat{S}}_{f,x'} \cdot \ve{\hat{S}}_{f,x'+1} \rangle. 
\end{equation}

\subsubsection{Bond inversion} 
\begin{figure}[h]
	\begin{center}
		\includegraphics[width=0.45\columnwidth,angle=-0]{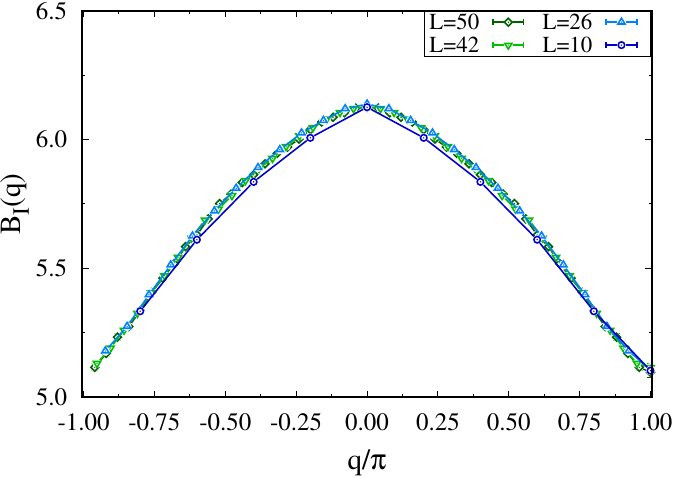}
		\caption{$B_I(q)$ as defined in Eq.~\ref{eq:BIq} as a function of the Heisenberg chain length $L$ in the VBS
			phase at $J_H/J_K=0.8$.  QMC parameters as in Fig.~\ref{fig:channel_symm}. }
		\label{fig:Bond_inversion}           
	\end{center}
\end{figure}

Under bond inversion, $\hat{V}^{}_{n,x}$ maps onto $\hat{V}^{}_{n,-x}$.  Charge density waves break bond inversion symmetry  whereas a valence bond solid does not.  
To  check  for  bond inversion symmetry  breaking,  we hence consider the  quantity:  
\begin{equation}
	\hat{O}_{x}    = \sum_{i=1}^{K}\hat{V}^{\dagger}_{i}(x) \hat{V}^{\phantom\dagger}_{i}(x) 
\end{equation}
and compute the correlation function 
\begin{equation}
	B_I(q) =  \sum_{x} e^{iqx}  \left( \langle \hat{O}_{x}\hat{O}_{x'} \rangle  - 
	\langle \hat{O}_{x} \rangle \langle \hat{O}_{x'} \rangle   \right).
	\label{eq:BIq}
\end{equation}
In all our simulations, we have never detected  bond inversion symmetry breaking, see Fig.~\ref{fig:Bond_inversion}.

\subsubsection{U(1) charge conservation symmetry of bath electrons  }
\begin{figure}[h]
	\begin{center}
		\includegraphics[width=0.45\columnwidth,angle=-0]{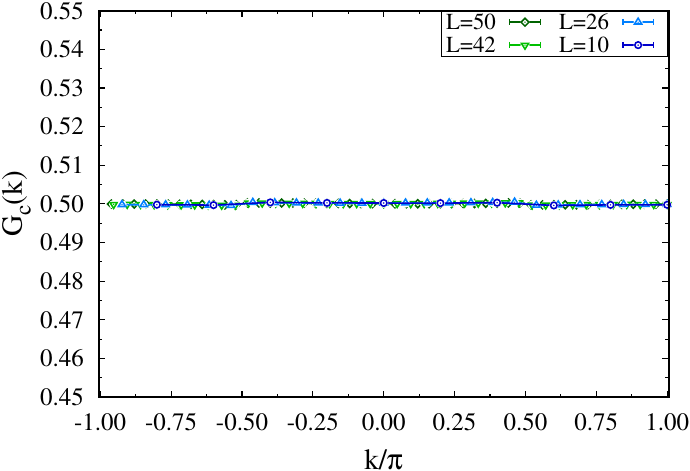}
		\caption{ Equal  time Green function $G_c(k)$  as defined in Eq.~\ref{eq:Greenc} as a function of the 
			Heisenberg chain length $L$ in the VBS phase at $J_H/J_K=0.8$. QMC parameters as in Fig.~\ref{fig:channel_symm}.  
			The very small deviations from 0.5   are a consequence of the finite Trotter time step. }
		\label{fig:Greenc}           
	\end{center}
\end{figure}

Our  Hamiltonian commutes with $  \hat{N}_x = \sum_{i,\alpha,y} \hat{c}^{\dagger}_{i\alpha}(x,y)  
\hat{c}^{\phantom\dagger}_{i\alpha}(x,y)$.  
It is interesting to ask the question if single bath electrons can propagate along the chain in the Kondo state.  To detect this,  we  have computed  the equal  time Green function
\begin{equation}
	G_c(k)  =  \frac{1}{NK}\sum_{x} \sum_{i,\alpha}e^{i qx } \langle \hat{c}^{}_{i\alpha}(x,y=0) \hat{c}^{\dagger}_{i\alpha}(0,y=0)\rangle.
	\label{eq:Greenc}
\end{equation}
Our results in Fig.~\ref{fig:Greenc} show that in the Kondo  phase,  a single particle  hopping process  between different baths is blocked since  $G_c(k)$  is flat in 
momentum space. 

\begin{figure}[h]   
	\begin{center}
		\includegraphics[width=0.4\columnwidth,angle=-0]{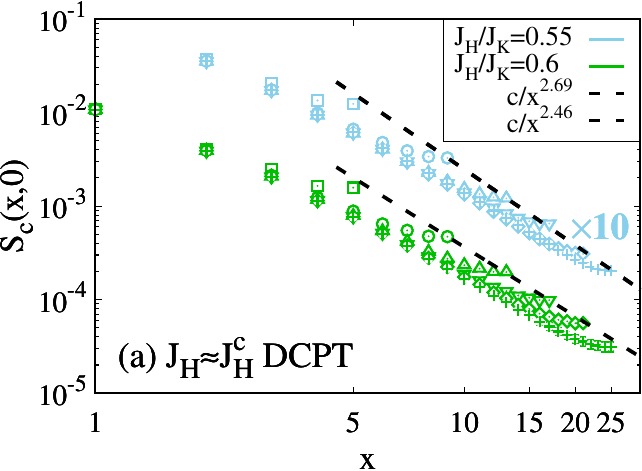} 
		\includegraphics[width=0.4\columnwidth,angle=-0]{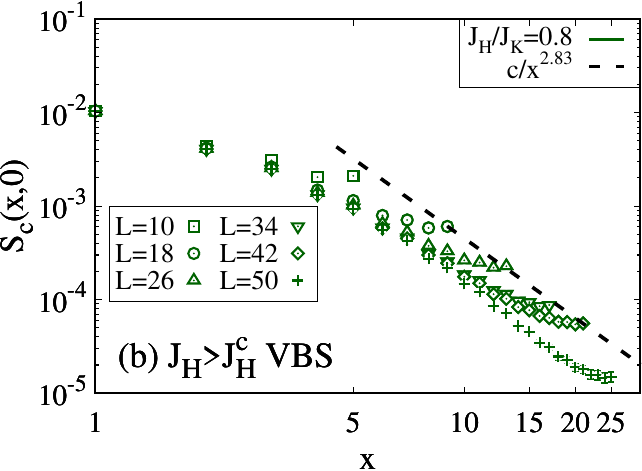}
		\caption{ Equal  time spin-spin correlation function $S_c(x,0)$ for the bath electrons: 
			(a) around  the 'dimension-changing' phase transition (DCPT) at $J_H^c/J_K^{}\simeq 0.55$ and 
			(b) in the VBS phase. }
		\label{fig:Spinc}           
	\end{center}
\end{figure}

On the other hand, the bath electron spin-spin correlations $S_c(x,0)$  along the chain are non-vanishing and within the errorbars decay with 
the same exponents as those of the  SU($N$)  spin  1/2  degrees of freedom, see Fig.~\ref{fig:Spinc}.

\subsection{Dynamical spin structure factor: generation of dispersion}

\begin{figure}[h]
	\begin{center}
		\includegraphics[width=0.4\columnwidth,angle=-0]{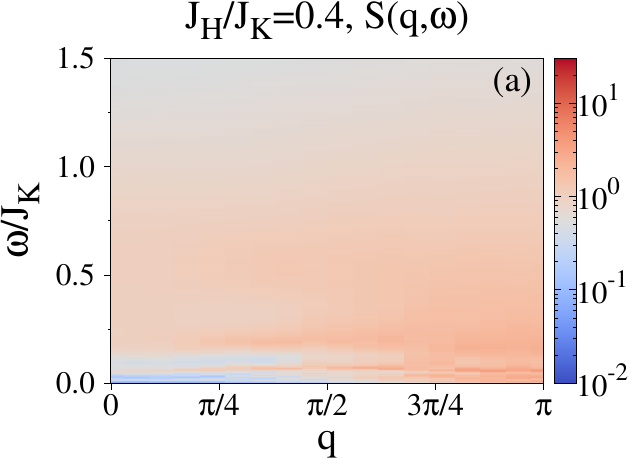}
		\includegraphics[width=0.4\columnwidth,angle=-0]{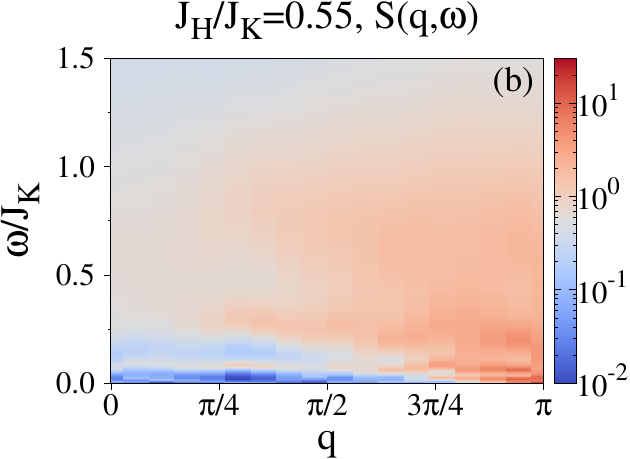}\\
		\includegraphics[width=0.4\columnwidth,angle=-0]{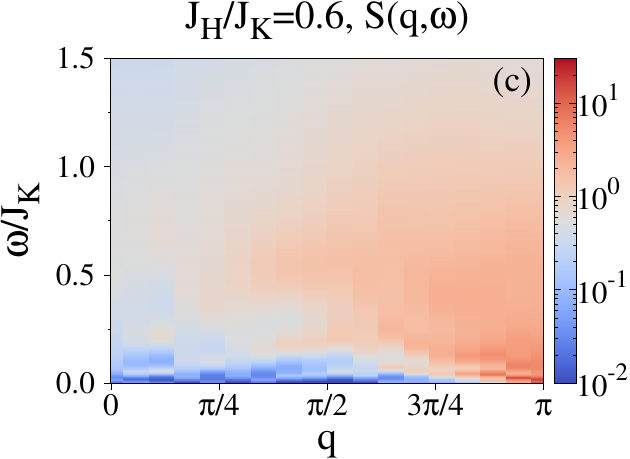}
		\includegraphics[width=0.4\columnwidth,angle=-0]{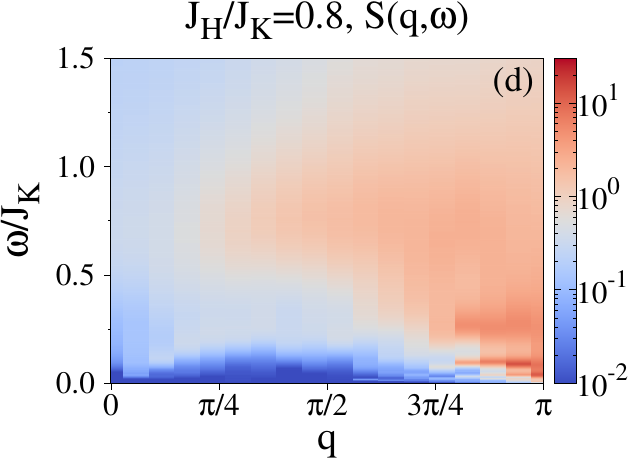}\\
		\includegraphics[width=0.4\columnwidth,angle=-0]{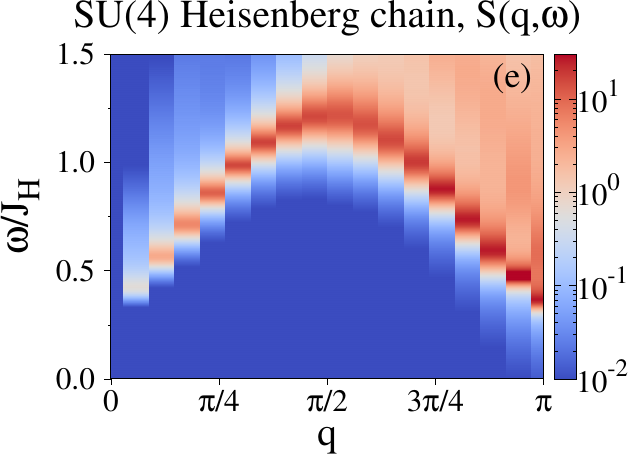}
		\caption{Dynamical spin structure factor  $S(q,\omega)$: 
			(a) in the decoupled phase at $J_H/J_K=0.4$,
			(b,c) around the DCPT at $J_H^{c}/J_K^{}\simeq 0.55$, and 
			(d) in the VBS phase at $J_H/J_K=0.8$. For comparison, panel (e) shows $S(q,\omega)$ of the decoupled ($J_K=0$) 
			SU(4) Heisenberg chain. }
		\label{fig:Sqw}
	\end{center}
\end{figure}

The dimensionality-changing transition discussed in the main text has clear signatures in  the spin dynamics.  In particular  dispersion is generated across 
the transition. However, due the bath  spin excitations remain gapless.  Figure~\ref{fig:Sqw}   explicitly shows this.   We equally compare the data to  isolated  SU(4) chain,  with apparent spin gap  and two-spinon continuum  above it.

The dynamical spin structure factor $S(q,\omega)$ at zero temperature is computed using  the  ALF library \cite{ALF_v2}  implementation of the stochastic analytical continuation methods \cite{beach04identifying,sandvik1998stochastic} via the relation

\begin{equation}
	S(q,\tau) =  \frac{1}{\pi} \int d \omega e^{-\omega \tau} S(q,\omega) 
\end{equation}
with 
\begin{equation}
	S(q,\tau) = \sum_{x} e^{iq x }   \sum_{a} \langle  \hat{S}^{a}_{f}(x,\tau)  \hat{S}^{a}_{f}(x=0,\tau=0) \rangle 
\end{equation}
the time displaced spin-spin correlation function.

\subsection{Robustness of power law spin correlations $\sim 1/\tau^2$  in the VBS phase}

\begin{figure}[h!]
	\begin{center}
		\includegraphics*[width=0.4\columnwidth,angle=-0]{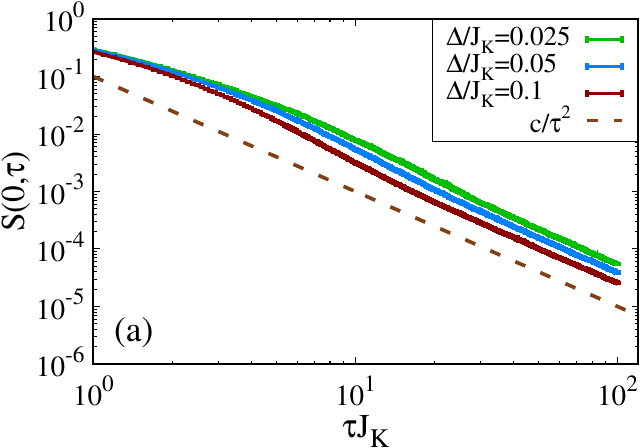}
		\includegraphics*[width=0.4\columnwidth,angle=-0]{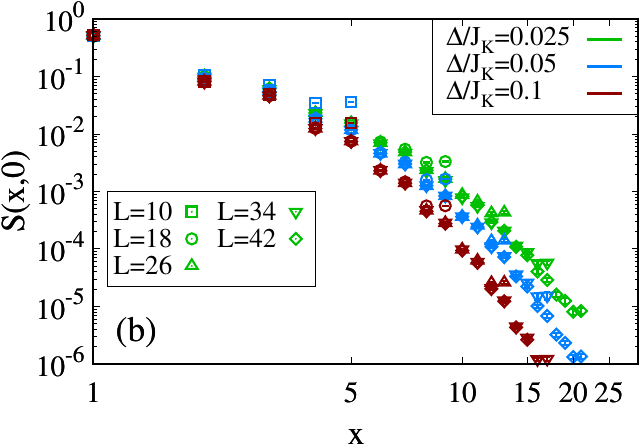}
		\caption{Space and time spin-spin correlation functions $S(x,\tau)$
			at $J_H/J_K=0.8$ in the VBS phase with a mass term $\Delta/J_K=0.025$, 0.05. and 0.1.
		}
		\label{fig:S_Delta}
	\end{center}
\end{figure}

We  check now the robustness of the power law decay of spin-spin correlations $S(x=0,\tau) \sim 1/\tau^2$  in the VBS phase. 
To this end, we modify our Hamiltonian by considering  a staggered Heisenberg interaction
\begin{equation}
	\hat{H}_{J_H}= \frac{2}{N} \sum_{x=1}^{L} \sum_{a=1}^{N^2-1} \bigl[ J_H + (-1)^{x}\Delta\bigr ] \hat{S}^a(x) \hat{S}^a(x+1)
	\label{eq:S_Delta}
\end{equation}
with a mass term $\Delta$. 
Figure~\ref{fig:S_Delta} (a) shows that $S(x=0,\tau)$ retains in the presence of finite $\Delta$ its power law tail at long times. 
On the other hand, equal time real space correlations $S(x,\tau=0)$ decay exponentially as in an isolated one-dimensional chain with VBS order, 
see Fig.~\ref{fig:S_Delta} (b).

\section{Mean-field for the VBS phase}\label{Sec:Appendix_C}

\subsection{Spin-spin correlations  using low-energy effective action}

In a standalone VBS phase, the ground state lives in the spin-singlet subspace, while the spinful excitations are gapped. The low-energy action for the spinful excitations is therefore of the form $S_0 = \int d\omega dk\,\, |\vec{\phi}|^2 (\omega^2+k^2 + m^2)$, where $m$ is the mass to the spinful excitations. In the presence of Kondo coupling to the electronic bath, however, the spinful excitations are Landau-damped, leading to $S = \int d\omega dk\,\, |\vec{\phi}|^2 (k^2 + m^2 + \gamma |\omega| + ...)$, with $\gamma \propto J^2_K$ and where $J_K$ is the Kondo coupling. The ellipsis denotes more irrelevant terms. The unequal-time, equal-space spin-spin correlator $S(x=0,\tau)$ is therefore proportional to 

\begin{equation}
	S(x=0,\tau) \propto \int\frac{ dk d\omega \, e^{i \omega \tau}}{k^2 + m^2 + \gamma |\omega|}
	\sim \int\frac{d\omega \, e^{i \omega \tau}}{\sqrt{m^2 + \gamma |\omega|}}
	\sim \int\frac{d\omega \, e^{i \omega \tau} \gamma |\omega|}{m^3} \sim \frac{\gamma}{m^3\tau^2} \, .
\end{equation}

\noindent Note that the power-law arises since the integrand is non-analytic due to the Landau-damping term $ \gamma |\omega|$. Similarly, the equal-time, unequal-space correlator is proportional to

\begin{equation}
	S(x,\tau=0) \propto \int\frac{ dk d\omega \, e^{i k x}}{k^2 + m^2 + \gamma |\omega|} \sim \int dk e^{i k x} \log(1 + \frac{\Omega}{k^2 + m^2}) \sim \frac{\e^{-m|x|} - \e^{-\sqrt{m^2+\Omega^2}}}{|x|} \sim \frac{\e^{-m|x|}}{|x|} \, ,
\end{equation}

\noindent where $\Omega$ is the UV cutoff for the frequency integral. The integral decays exponentially with $|x|$ as expected, since the integrand is an analytic function of $k$.

\subsection{Large-$N$ slave-parton mean-field}
We will here provide a mean-field description of the VBS phase, where the mean-field parameters are taken to be homogeneous and time-independent. We will primarily focus on the spinon Green's function $G_f(x,\tau) \sim \langle T_\tau \hat{f}(x,\tau)  \hat{f}^{\dagger}(0,0)\rangle$, which, at the mean-field level, allows one to calculate the spin-spin correlations $S(x,\tau) = \langle \hat{S}^a(x,\tau) \hat{S}^a(0,0) \rangle \sim \left(G_f(x,\tau)\right)^2$. The main outcome of this mean-field is that equal-space, unequal-time correlations decay as a power-law, $S(x=0, \tau) \sim 1/\tau^2$, while equal-time, unequal-space correlations decay exponentially, $S(x, \tau = 0) \sim e^{-x/\xi_{VBS}}$, in line with the aforementioned conclusion based on the low-energy theory.

We start with our Hamiltonian (see below Eq. \ref{Eq:Hamiltonian} for more details)


\begin{align}
	\begin{split}
		\hat{H} &= 
		\sum_k \sum_{x=1}^L \epsilon(k) \hat{c}^{\dag}_{i \alpha}(x,k) \hat{c}_{i \alpha}(x,k) + \frac{J_K}{N}  \sum_x \hat{S}^a(x) \hat{c}^{\dag}_{i\alpha}(x,y=0) T^{a}_{\alpha \beta} \hat{c}_{i\beta}(x,y=0) \\ &\hspace{0.5cm} + \frac{J_H}{N} \sum_x \hat{S}^a(x) \hat{S}^a(x+1) \, ,
	\end{split}
\end{align}

\noindent where the sums over $\alpha$ (spin), $i$ (channel) and $a$ (spin components) are implicit and where the couplings have been rescaled by $1/N$. We will also denote $\hat{c}_{i\alpha}(x) \equiv \hat{c}_{i\alpha}(x,y=0)$ for now on. The local moments are written in terms of fermionic spinons $\hat{f}_{\alpha}(x)$

\begin{equation}
	\hat{S}^a(x) = \hat{f}_{\alpha}^{\dag}(x) T_{\alpha \beta}^a \hat{f}_{\beta}(x) , \quad \sum_{\alpha = 1}^N \hat{f}_{\alpha }^{\dag}(x) \hat{f}_{\alpha}(x) = \frac{N}{2} \, .
\end{equation}

\noindent This results in two quartic terms in the Hamiltonian, which can be decoupled by introducing two Hubbard-Stratonovich fields $\hat{\phi}_{i}^{\dag}(x) \sim \sum_{\alpha} \hat{f}_{\alpha }^{\dag}(x) \hat{c}_{i\alpha}(x)$ and $\hat{\chi}^{\dag}(x) \sim \sum_{\alpha} \hat{f}_{\alpha}^{\dag}(x) \hat{f}_{\alpha}(x+1)$. By dropping an unimportant constant and a shift to the conduction electron chemical potential, the Hamiltonian becomes

\begin{align} \label{Eq:H_decoupled}
	\begin{split}
		\hat{H} &= \sum_{k} \sum_{x} \epsilon(k) \hat{c}^{\dag}_{i \alpha}(x,k) \hat{c}_{i \alpha}(x,k) - \sum_{x} \Big[ \hat{\phi}_{i}^{\dag}(x) \hat{c}_{i \alpha}^{\dag}(x) \hat{f}_{\alpha}(x) + \hat{f}_{\alpha}^{\dag}(x) \hat{c}_{i\alpha}(x) \hat{\phi}_{i}(x) \Big] \\ &\hspace{0.5cm} - \sum_{x} \Big[ \hat{\chi}^{\dag}(x) \hat{f}_{\alpha}^{\dag}(x+1) \hat{f}_{\alpha}(x) + \hat{f}_{\alpha}^{\dag}(x) \hat{f}_{\alpha}(x+1) \hat{\chi}(x) \Big]  \\ &\hspace{0.5cm} + \frac{N}{J_K} \sum_{x} \hat{\phi}_{i}^{\dag}(x) \hat{\phi}_{i}(x) + \frac{N}{J_H} \sum_x \hat{\chi}^{\dag}(x) \hat{\chi}(x) \, .
	\end{split}
\end{align}

The mean-field Hamiltonian is obtained by starting with Eq. \ref{Eq:H_decoupled} and by condensing $\hat{\phi}_i(x) = \hat{\phi}_i^{\dag}(x) = \phi$ to a constant. $\hat{\chi}$ is also condensed, but due to dimerization takes the form

\begin{equation}
	\hat{\chi}(x) = \hat{\chi}^{\dag}(x) = \begin{cases} \chi_A = \chi_0(1-\delta)\text{ for } x \in A \text{ sublattice} \\ \chi_B = \chi_0(1+\delta) \text{ for } x \in B \text{ sublattice} \end{cases} \, ,
\end{equation}

\noindent where $A$ and $B$ sublattices respectively correspond to odd and even sites. Defining the conduction electron and spinon operators on the two sublattices via their Fourier transform in the reduced Brillouin zone,

\begin{align}
	\begin{split}
		\hat{c}_{i \alpha}^{A/B}(x',y) = \sqrt{\frac{2}{L}} \sum_{k_x} \e^{2i k_x x'} \hat{c}_{i \alpha}^{A/B}(k_x,y) \, , \quad \hat{f}_{\alpha}^{A/B}(x') = \sqrt{\frac{2}{L}} \sum_{k_x} \e^{2i k_x x'} \hat{f}_{\alpha}^{A/B}(k_x) \, , 
	\end{split}
\end{align}

\noindent where $x' = 1,..., L/2$ labels the new unit cell and $k_x \in ]-\pi/2,\pi/2]$, the mean-field Hamiltonian becomes

\begin{align}
	\begin{split}
		\hat{H} &= \sum_{k_x,k_y} \epsilon(k_y) \Big[ \hat{c}_{i\alpha}^{A \dag}(k_x,k_y) \hat{c}_{i\alpha}^A(k_x,k_y) + \hat{c}_{i\alpha}^{B \dag}(k_x,k_y) \hat{c}_{i\alpha}^B(k_x,k_y) \Big] \\ &\hspace{0.5cm}-\phi \sum_{k_x} \Big[ \hat{f}_{\alpha}^{A \dag}(k_x) \hat{c}^A_{i\alpha}(k_x) + \hat{f}_{\alpha}^{B \dag}(k_x) \hat{c}^B_{i\alpha}(k_x) + h.c. \Big] \\ &\hspace{0.5cm} - \sum_{k_x} \Big[ \Big( \chi_A + \chi_B \e^{-2i k_x} \Big) \hat{f}_{\alpha}^{A \dag}(k_x) \hat{f}_{\alpha}^B(k_x) + h.c. \Big]  + \frac{N}{J_K} L \phi^2 + \frac{N}{J_H} \frac{L_y}{2} \big( \chi_A^2 + \chi_B^2 \big) \, ,
	\end{split}
\end{align}

\noindent where the bath momentum is now denoted $k_y$ to avoid confusion. We are now in a position to compute the spinon 2-point function, which can be done analytically. We can drop the last two constant terms as they do not affect the correlation function calculation. We start by obtaining the effective action for $f$ by integrating out the conduction electrons. To do so, one must work with the imaginary-time action. Transforming to Matsubara frequencies, one finds

\begin{align}
	\begin{split}
		S &= \sum_{\Vec{k}} \sum_{\omega_n} \sum_{i, \alpha} \bar{\psi}_{i \alpha}(\Vec{k},i \omega_n) \begin{pmatrix}
			-i \omega_n +\epsilon(k_y) & 0 \\ 0 & -i \omega_n + \epsilon(k_y)
		\end{pmatrix} \psi_{i \alpha}(\Vec{k},i \omega_n) \\ &-\frac{\phi}{\sqrt{L_b}} \sum_{\Vec{k}} \sum_{\omega_n} \sum_{i, \alpha} \Bigg[ \bar{\psi}_{i \alpha}(\Vec{k},i \omega_n) \eta_{\alpha}(k_x,i \omega_n) + \bar{\eta}_{\alpha}(k_x,i \omega_n) \psi_{i \alpha}(\Vec{k},i \omega_n) \Bigg] \\ &+ \sum_{k_x} \sum_{\omega_n} \sum_{\alpha} \bar{\eta}_{\alpha}(k_x,\omega_n) \begin{pmatrix}
			-i \omega_n & - z(z_k) \\ - z^*(k_x) & -i \omega_n
		\end{pmatrix} \eta_{\alpha}(k_x,i \omega_n) \, ,
	\end{split}
\end{align}

\noindent where $\Vec{k} = (k_x,k_y)$, $z(k_x) = \chi_A + \chi_B \e^{-2i k_x}$ and the two following 2-component spinors have been introduced

\begin{equation}
	\psi_{i \alpha}(\Vec{k},i \omega_n) = \begin{pmatrix}
		c^A_{i \alpha}(\Vec{k},i \omega_n) \\ c^B_{i \alpha}(\Vec{k},i \omega_n)
	\end{pmatrix} \, , \quad \eta_{\alpha}(k_x,i \omega_n) = \begin{pmatrix}
		f^A_{\alpha}(k_x,i \omega_n) \\ f^B_{\alpha}(k_x,i \omega_n)
	\end{pmatrix} \, .
\end{equation}

\noindent Note that the local conduction electron operator has also been written in momentum space. We then define the following free Green's functions

\begin{align}
	\begin{split}
		-G_{c,0}^{-1}(k_y,i \omega_n) &= \begin{pmatrix}
			-i \omega_n +\epsilon(k_y) & 0 \\ 0 & -i \omega_n + \epsilon(k_y)
		\end{pmatrix} \\ -G_{f,0}^{-1}(k_x,i \omega_n) &= \begin{pmatrix}
			-i \omega_n & - z(z_k) \\ -z^*(k_x) & -i \omega_n
		\end{pmatrix} \, .
	\end{split}
\end{align}

Conduction electrons can now be integrated out leading to the following effective action for $f$:
\begin{align}
	\begin{split}
		S_{\text{Eff}}[\eta] = \sum_{k_x} \sum_{\omega_n} \bar{\eta}_{\alpha}(k_x,i \omega_n) \Big[ -G_{f,0}^{-1}(k_x,i \omega_n) + K \phi^2 \Big( \frac{1}{L_b} \sum_{k_y} G_{c,0}(k_y,i \omega_n) \Big) \Big] \eta_{\alpha}(k_x,i \omega_n) \, .
	\end{split}
\end{align}

\noindent From this, the (inverse) propagator for $f$ is identified

\begin{align}
	\begin{split}
		G_f^{-1}(k_x,i \omega_n) &= - \Bigg[ -G_{f,0}^{-1}(k_x,i \omega_n) + K \phi^2 \Big( \frac{1}{L_b} \sum_{k_y} G_{c,0}(k_y,i \omega_n) \Big) \Bigg] \\ &= \begin{pmatrix}
			i \omega_n - \frac{1}{L_b} \sum_{k_y} \frac{K \phi^2}{i \omega_n - \epsilon(k_y)} & z(k_x) \\ z^*(k_x) & i \omega_n - \frac{1}{L_b} \sum_{k_y} \frac{K \phi^2}{i \omega_n - \epsilon(k_y)} 
		\end{pmatrix} \, .
	\end{split}
\end{align}

\noindent The sum over $k_y$ is performed using a constant density of states at the Fermi level and a bandwidth $D$

\begin{align}
	\begin{split}
		\frac{1}{L_b} \sum_{k_y} \frac{1}{i \omega_n - \epsilon(k_y)} &\approx d(0) \int_{-D}^D d\epsilon \frac{1}{i \omega_n - \epsilon} \\ &= d(0) \int_{-D}^{D} d\epsilon \frac{-i \omega_n - \epsilon}{\omega_n^2 + \epsilon^2} \\ &= -2i d(0) \arctan\Big(\frac{D}{\omega_n}\Big) \\ &\approx -i \pi d(0) \sgn(\omega_n) \, ,
	\end{split}
\end{align}

\noindent where the last equality holds for $D \gg \omega_n$. Inverting the matrix and defining the constant $C_K = \pi d(0) K \phi^2$ yields

\begin{align}
	\begin{split}
		G_f(k_x,i \omega_n) = \frac{-1}{\big( \omega_n + C_K \sgn(\omega_n)\big)^2 + |z(k_x)|^2 \big)} \begin{pmatrix}
			i \omega_n + i C_K \sgn(\omega_n) & -z(k_x) \\ -z^*(k_x) & i \omega_n + i C_K \sgn(\omega_n)
		\end{pmatrix} \, .
	\end{split}
\end{align}

\noindent We can now compute the correlation function by Fourier transforming the above expression. We work at $T=0$ for now on. 

The equal-space, unequal-time spinon correlation function is computed by working on the same sublattice ($AA$ here). Hence

\begin{align}
	\begin{split}
		G_f(x=0,\tau) &= \frac{2}{L} \sum_{k_x} \int \frac{d\omega}{2\pi} \e^{-i \omega \tau} [G_F(k_x,i \omega)]_{AA} \\ &\approx -\int_{-\pi/2}^{\pi/2} \frac{dk_x}{2\pi} \int_{-\infty}^{\infty} \frac{d\omega}{2\pi} \e^{-i \omega \tau} \frac{i \omega + i C_K \sgn(\omega)}{(\omega+C_K \sgn(\omega))^2 + |z(k_x)|^2} \\ &= -2 \int_{-\pi/2}^{\pi/2} \frac{dk_x}{2\pi} \int_{0}^{\infty} \frac{d\omega}{2\pi} \sin(\omega \tau) \frac{\omega + C_K}{(\omega+C_K)^2 + |z(k_x)|^2} \\ &\overset{\tau \to \infty}{\sim} \frac{1}{\tau} \, ,
	\end{split}
\end{align}

\noindent where the last line can be shown by asymptotically performing the integral or computing it numerically. This result holds for any value of $\delta$. Hence, this confirms that in the VBS phase, the equal-space, unequal-time spin-spin correlation function decays as $1/\tau^2$ ($S(x=0,\tau) \sim 1/\tau^2$).

We now move on to the equal-time, unequal-space correlation function. This time, one must take the off-diagonal elements of the matrix (different sublattice, $AB$ here). Therefore

\begin{align}
	\begin{split}
		G_f(x,\tau=0) &= \frac{2}{L} \sum_{k_x} \e^{i k_x x} \int_{-\infty}^{\infty} \frac{d\omega}{2\pi} [G_f(k_x,i \omega)]_{AB} \\ &\approx \int_{-\pi/2}^{\pi/2} \frac{dk_x}{2\pi} \e^{2i k_x x} \int_{-\infty}^{\infty} \frac{d\omega}{2\pi} \frac{z(k_x)}{(\omega + C_K \sgn(\omega))^2 + |z(k_x)|^2} \\ &= 2 \int_{-\pi/2}^{\pi/2} \frac{dk_x}{2\pi} \e^{2i k_x x} \int_{0}^{\infty} \frac{d\omega}{2\pi}\frac{z(k_x)}{(\omega+C_K)^2 + |z(k_x)|^2} \\ &= \frac{1}{4\pi^2} \int_{-\pi/2}^{\pi/2} dk_x \e^{2i k_x x} \Big( \pi - 2 \arctan(\frac{C_K}{|z(k_x)|}) \Big) \frac{z(k_x)}{|z(k_x)|} \, ,
	\end{split}
\end{align}

\noindent where the integral over $\omega$ was performed exactly in the last step. The integral over $k_x$ corresponds to the Fourier transform of an analytic function since $|z(k_x)| > 0$, and therefore $G_f(x,\tau=0) \sim e^{- x/\xi}$ where $\xi$ is non-universal and finite, highlighting the exponential decay of $S(x,\tau=0)$.

\end{document}